\documentclass[preprints,article,accept,moreauthors,pdftex]{Definitions/mdpi}

\firstpage{1}
\pubvolume{1}
\issuenum{1}
\articlenumber{0}
\pubyear{2021}
\copyrightyear{2021}
\externaleditor{Academic Editor: Francesca Loi and Tiziana Venturi} 
\datereceived{30 September 2021}
\dateaccepted{27 October 2021}
\datepublished{}
\hreflink{https://doi.org/} 

\usepackage{amssymb}
\usepackage[labelformat=simple]{subcaption}

\DeclareCaptionLabelFormat{subcaptionlabel}{\normalfont(\textbf{#2}\normalfont)}
\newcommand\arcsec{\hbox{$^{\prime\prime}$}}

\Title{Exploring New Redshift Indicators for Radio-Powerful AGN}

\TitleCitation{Exploring New Redshift Indicators for Radio-Powerful AGN}

\Author{Rodrigo Carvajal $^{1,2,}$*\orcidA{}, Israel Matute $^{1,2}$\orcidB{}, José Afonso $^{1,2}$\orcidC{}, Stergios Amarantidis $^{1,2}$\orcidD{}, Davi Barbosa $^{1,2}$\orcidE{}, Pedro~Cunha~$^{3,4}$\orcidF{} and Andrew Humphrey $^{3}$\orcidG{}}

\AuthorNames{Rodrigo Carvajal, Israel Matute, José Afonso, Stergios Amarantidis, Davi Barbosa, Pedro Cunha and Andrew Humphrey}

\AuthorCitation{Carvajal, R.; Matute, I.; Afonso, J.; Amarantidis, S.; Barbosa, D.; Cunha, P.; Humphrey, A.}

\address{
$^{1}$ \quad Instituto de Astrofísica e Ciências do Espaço, Universidade de Lisboa, OAL, Tapada da Ajuda, \mbox{PT1349-018 Lisbon}, Portugal; imatute@oal.ul.pt (I.M.); jafonso@iastro.pt (J.A.); samarant@oal.ul.pt (S.A.); dbarbosa@oal.ul.pt (D.B.)\\
$^{2}$ \quad Departamento de Física, Faculdade de Ciências, Universidade de Lisboa, Edifício C8, Campo Grande, \mbox{PT1749-016 Lisbon}, Portugal\\
$^{3}$ \quad Instituto de Astrofísica e Ciências do Espaço, Universidade do Porto, CAUP, Rua das Estrelas, \mbox{PT4150-762 Porto}, Portugal;
pedro.cunha@astro.up.pt (P.C.); andrew.humphrey@astro.up.pt (A.H.)\\
$^{4}$ \quad Departamento de Física e Astronomia, Faculdade de Ciências, Universidade do Porto, Rua do Campo Alegre~687, \mbox{PT4169-007 Porto}, Portugal}

\corres{Correspondence: rcarvajal@oal.ul.pt (R.C.)}

\abstract{Active Galactic Nuclei (AGN) are relevant sources of radiation that might have helped reionising the Universe during its early epochs. The super-massive black holes (SMBHs) they host helped accreting material and emitting large amounts of energy into the medium. Recent studies have shown that, for epochs earlier than \texorpdfstring{$z~{\sim}~5$}{z ~ 5}, the number density of SMBHs is on the order of few hundreds per square degree. Latest observations place this value below $300$ SMBHs at \texorpdfstring{$z~{\gtrsim}~6$}{z ≳ 6} for the full sky. To overcome this gap, it is necessary to detect large numbers of sources at the earliest epochs. Given the large areas needed to detect such quantities, using traditional redshift determination techniques---spectroscopic and photometric redshift---is no longer an efficient task. Machine Learning (ML) might help obtaining precise redshift for large samples in a fraction of the time used by other methods. We have developed and implemented an ML model which can predict redshift values for WISE-detected AGN in the HETDEX Spring Field. We obtained a median prediction error of \texorpdfstring{$\sigma_{z}^{N} = 1.48 \times (z_{\mathrm{Predicted}} - z_{{\mathrm{True}}}) / (1 + z_{\mathrm{True}}) = 0.1162$}{σ_zN = 1.48 * (z_Predicted - z_True) / (1 + z_True) = 0.1162} and an outlier fraction of \texorpdfstring{$\eta = 11.58 \%$}{η = 11.58 \%} at \texorpdfstring{$(z_{\mathrm{Predicted}} - z_{{\mathrm{True}}}) / (1 + z_{\mathrm{True}}) > 0.15$}{(z_Predicted - z_True) / (1 + z_True) > 0.15}, in line with previous applications of ML to AGN. We also applied the model to data from the Stripe 82 area obtaining a prediction error of \texorpdfstring{$\sigma_{z}^{N} = 0.2501$}{σ_zN = 0.2501}.}



\begin{document}


\section{Introduction}\label{sec:intro}

Super-Massive Black Holes (SMBHs) might be ubiquitous to all galaxies above a certain mass. Understanding their true role in the shaping of galaxies will require a more precise census of the nature, growth, and evolution of SMBHs---in the so-called Active Galactic Nuclei (AGN) phases---, as well as a more detailed characterisation of the internal (secular) and external (environment) processes at work within the host~\cite{2017A&ARv..25....2P}.

Radio selection has been traditionally a prime wavelength for the detection of AGN activity.
Between 10--20\% of AGN have strong radio emission, in~many cases in the form of jets, that can overshadow the radio emission associated to star-forming regions, mostly due to super-novae \cite{2014ARA&A..52..589H}. Radio selection efficiency though seems to decrease towards the Epoch of Reionisation (EoR. $z~{>}~6$, e.g., Reference \cite[]{2006ApJ...652..157M, 2021ApJS..253...25K, 2021MNRAS.501.3833D, 2021ApJ...915..126L}). Simulations (e.g., Reference \cite[]{2019MNRAS.485.2694A, 2021MNRAS.503.3492T, 2019MNRAS.482....2B}) predict that the distribution of AGN and Radio Galaxies (RG) along redshift can lead to the detection of a few hundreds of objects per square degree at the EoR as the with deep observations planned for future observatories, e.g., SKA, $\mu$Jy sensitivity levels~\cite{2015aska.confE..67P}. These expectations collide with the most recent compilations (see, for~instance, Reference \cite[]{2020ARA&A..58...27I, 2020MNRAS.494..789R}), which show that only ${\sim}300$ sources have been confirmed to exist at $z~{>}~6$ over most of the sky. Environmental (CMB) and intrinsic (QSO versus radio mode accretion) conditions might be responsible for the lower rate of radio powerful sources at $z~{>}~5$ but, selection criteria, might also be playing a role~\cite{2008A&ARv..15...67M}. Nevertheless, current radio instruments and recently completed surveys~\cite{2015ApJ...801...26H, 2011PASA...28..215N, 2020RNAAS...4..175G, 2019A&A...622A...1S} have allowed detection of larger numbers of RG (e.g., Reference \cite[]{2014A&A...569A..52S, 2018MNRAS.475.3429W, 2020A&A...642A.107C}) that could be used to better understand the origin and evolution of radio emission in~AGN.

To place radio AGN in the proper cosmological context and derive their intrinsic properties, and~given the time constrains imposed for the compilation of significant spectroscopic samples, alternative estimates for redshift need to be used.
Template-based photometric redshifts have proven to be an efficient alternative by trading precision for sky coverage. The~sizes of the new catalogues though, with~tens to hundred of millions of sources, imply a significant---and ever increasing---investment in computational time. These issues raise the need for additional approaches which might be able to obtain the redshift information for a large number of astrophysical sources with enough precision and within a reasonable amount of~time.

The tremendous increase of computing power over the last decades has allowed the application of evolved statistical methods in the analysis of large and complex datasets. Using previously-fed data, it is possible to predict, with~relevant confidence, the~behaviour new data will have. This is what has been called Machine Learning (ML). In~Astrophysics, ML has been used in a wide range of subjects (in AGN and other types of sources), such as redshift determination (e.g., Reference \cite[]{2021A&A...649A..81N, 2021AJ....162...72W}), morphological classification (e.g., Reference \cite{2019ApJS..240...34M, 2019MNRAS.487.1729L, 2021A&A...645A..89M, 2021A&A...648A.102V, 2021MNRAS.505.4345B}), source selection and classification (e.g., Reference \cite{2016ApJ...820....8S, 2016MNRAS.462.3180C, 2020A&C....3200387X, 2021arXiv210612787W}), image and spectral reconstruction (e.g., Reference \cite{2021PNAS..11822038L}), and more \cite{2010IJMPD..19.1049B, 2019arXiv190407248B}. Despite its range of applications, ML has received some criticism related to the interpretability of the derived models, e.g.,~most ML models cannot provide coefficients that allow to create an analytical expression for example~\cite{goebel2018explainable}. This implies that it may not be straightforward to understand the exact role that the measured properties have into the prediction a model might~make.

Recent work has been done to improve interpretability. Feature importance~\cite{9007737} can be derived, mostly for Tree-Based models, i.e., models that use decision trees to classify or predict properties. In~this scenario, a~feature with a high importance will be, in~general, in~the higher levels of the decision trees used for the modelling.
A different method for assessing the impact of features is that of Shapley Values~\cite{Shapley_article}. Opposite to feature importance, Shapley values, which have been defined in the context of Game Theory to determine the contribution of a player in a cooperative game, can help in understanding how the features impact each individual prediction. A~more thorough description on how Shapley values work can be seen in \citet{molnar2019}.

Astronomical data is very heterogeneous in its current form, with~small areas of the sky covered extensively at all wavelengths and with high sensitivity but also larger areas with sparser multiwavelength coverage. Therefore, the~homogeneous and deep multiwavelength coverage required for the most accurate models can only be achieved over a few to tens of degrees. The~derived models in these fields could then be applied first to present surveys with less extensive and deep multiwavelength coverage (e.g., LoTSS, Stripe82, RACS, MIGHTEE, etc.) and then also to the upcoming all-sky surveys, e.g., SKA, LSST, eROSITA, etc., delivering observations with comparable depth and multiwavelength coverage as current small~fields.

In this work, we describe an ML model aiming to predict the redshift for AGN based on the multiwavelength properties of the HETDEX Spring Field with the minimum amount of data preparation possible. The~model will be then tested in data from the SDSS Stripe 82 Field where multiwavelength coverage and depth vary with respect to the HETDEX Spring Field. This approach would test the validity of the derived ML model based on a field into other fields with slightly different spectral~coverage.

The structure of this article is as follows. In~Section~\ref{sec:Mat_meth}, we present the used data, its preparation for ML training and describe the selection of models and the metrics used for assess their results. In~Section~\ref{sec:results}, the~results of model training and validation are shown, as well as the predictions over the Stripe 82 field. We present the discussion of our results in Section~\ref{sec:discussion}. Finally, in~Section~\ref{sec:conclusions}, we summarise our~work.

\section{Materials and~Methods}\label{sec:Mat_meth}
\unskip

\subsection{Data}\label{sec:data}

\begin{figure}[H]
\includegraphics[width=0.8\columnwidth]{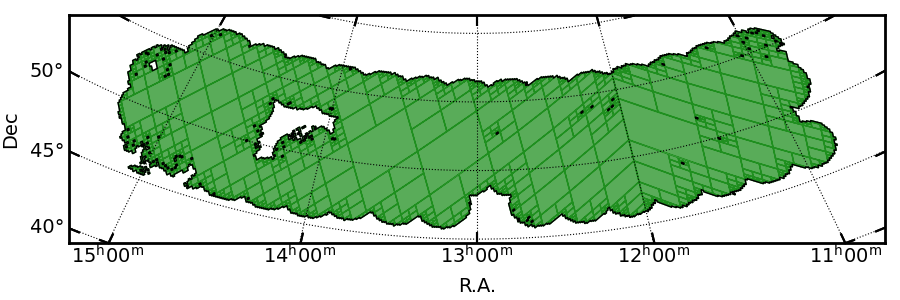}
\caption{Area of the HETDEX Spring Field covered by the LOFAR DR1 measurements. Figure prepared, in part, using the Python package \texttt{MOCPy} \cite{2014ivoa.spec.0602F}.\label{fig:hetdex_area}} 
\end{figure}

We have selected all the detections listed on the CatWISE2020 catalogue (CW, \cite{2021ApJS..253....8M}) that are located in the HETDEX Spring Field and that have been covered by the LOFAR DR1 measurements~\cite{2019A&A...622A...1S} (see Figure~\ref{fig:hetdex_area}). The~CatWISE2020 catalogue has measurements in the WISE bands W1 and W2 (at 3.4 $\upmu$m and 4.6 $\upmu$m, respectively), with~a $90\%$ completeness depth at W1 = 17.7 mag and W2 =17.5 mag. LOFAR observations cover an area of $424$~$\mathrm{deg}^{2}$, with~a median sensitivity of $71 \mu$Jy/beam and a $6 \arcsec$ resolution. In~that way, we have obtained 6,729,647 detected~sources.

The sources have been then cross-matched with other catalogues in different wavelengths using a search radius of $5 \arcsec$. We have selected surveys with large sky coverages, such as: VLASS (3 GHz) \cite{2020RNAAS...4..175G}, LoTSS-DR1 (150 MHz) \cite{2019A&A...622A...1S}, Pan-STARRS DR1~\cite{2020ApJS..251....7F, 2020ApJS..251....7F}, GALEX AIS GR6+7~\cite{2017ApJS..230...24B}, GMRT 150 MHz all-sky~\cite{2017A&A...598A..78I}, 4XMM-DR9~\cite{2020A&A...641A.137T}, 2MASS All-Sky~\cite{2003tmc..book.....C, 2006AJ....131.1163S}, and~AllWISE (AW \cite{2013wise.rept....1C}). The~$20$ selected photometric bands are listed in Table~\ref{tab:phot_bands}. To~homogenise photometric measurements, we converted all fluxes and magnitudes to AB~magnitudes.

\begin{specialtable}[H]
\caption{Photometric bands included in the~dataset.\label{tab:phot_bands}}

\setlength{\cellWidtha}{\columnwidth/4-2\tabcolsep+.0in}
\setlength{\cellWidthb}{\columnwidth/4-2\tabcolsep-0in}
\setlength{\cellWidthc}{\columnwidth/4-2\tabcolsep-0.0in}
\setlength{\cellWidthd}{\columnwidth/4-2\tabcolsep-0.0in}
\scalebox{1}[1]{\begin{tabularx}{\columnwidth}{>{\PreserveBackslash\centering}m{\cellWidtha}>{\PreserveBackslash\centering}m{\cellWidthb}>{\PreserveBackslash\centering}m{\cellWidthc}>{\PreserveBackslash\centering}m{\cellWidthd}}
\toprule
\textbf{Survey/Instrument}	& \textbf{Bands}  & \textbf{Survey/Instrument}	& \textbf{Bands}\\
\midrule
CatWISE2020 & W1, W2               & VLASS         & 3.0 GHz\\
AllWISE		& W1, W2, W3, W4       & GALEX         & FUV, NUV\\
Pan-STARRS  & g, r, i, z, y        & 2MASS         & J, H, K\\
LOFAR       & 150 MHz              & XMM-NEWTON    & 0.2--12 keV\\
GMRT        & 150 MHz              &               &   \\
\bottomrule
\end{tabularx}}
\end{specialtable}

\begin{figure}[H]
\widefigure
\begin{subfigure}[b]{0.45\columnwidth}
\includegraphics[width=\textwidth]{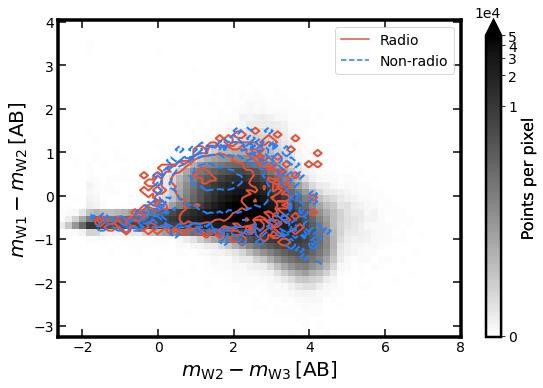}
\caption{WISE colour-colour~diagram.}
\end{subfigure}
\begin{subfigure}[b]{0.45\columnwidth}
\includegraphics[width=\textwidth]{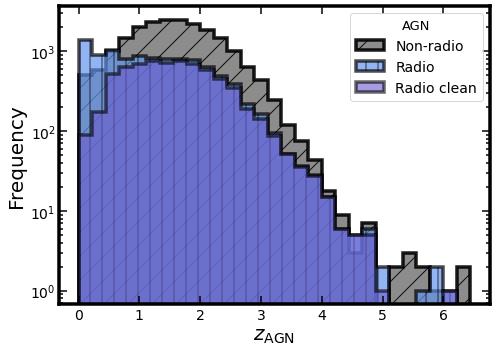}
\caption{Histograms for z~values.}
\end{subfigure}
\vspace{6pt}

\caption{Characterisation plots for the Active Galactic Nuclei (AGN) sources in the HETDEX Spring Field. (\textbf{a}) W1--W2, W2--W3 colour-colour diagram. Grey background represents the full CatWISE2020 sample, with~darker areas showing higher number of sources following the colour bar. Red, solid contours show density levels for radio-detected AGN. Blue, dashed contours indicate density levels for AGN without a counterpart on the radio surveys used in this work (i.e., without~radio detection). For~both contour plots, the~lines show the levels with $1$, $10$, $100$, and~$1000$ sources in each pixel. (\textbf{b})~Histograms for the redshift values of sources labelled as AGN. Grey, hatched histogram shows the distribution of redshifts for AGN without radio detections. Redshifts for all radio-detected AGN are presented by the blue, vertically-hatched histogram. Confirmed AGN without high host influence (see main text) that show a measurement on the surveys used in this work, are presented in purple, clean~histogram. }\label{fig:histograms_hetdex}
\end{figure}

We then selected the sources that could be linked to the emission of an AGN. Thus, we cross-matched our catalogue with the Million Quasar Catalog\endnote{\url{http://quasars.org/milliquas.htm} (accessed on 3 May 2021).}
 (MQC, v7.2, \cite{2021arXiv210512985F}). It lists published type-I QSOs/AGN, quasar candidates, type-II object and blazars along with the best available redshift values for each of them, i.e., spectroscopic or photometric redshifts. For~the HETDEX Spring Field, 32,365 objects have been identified, in~different surveys, as~AGN. That means that $0.48 \%$ of the detected CatWISE2020 sources have been identified as AGN, and~close to $8 \%$ (2674) of them are considered as QSO candidates. From~the identified AGN in our sample, 26,520 sources are listed in the Sloan Sky Digital Survey Quasar Catalogue DR16 (SDSS-Q DR16 \citep{2020ApJS..250....8L}), implying that the mean properties of the objects studied in this work are driven by the behaviour of SDSS QSOs. In~Figure~\ref{fig:histograms_hetdex}, we show the distribution of the sources in the WISE colour-colour diagram and the histogram of available redshifts.

\begin{figure}[H]
\widefigure
\includegraphics[width=.93\columnwidth]{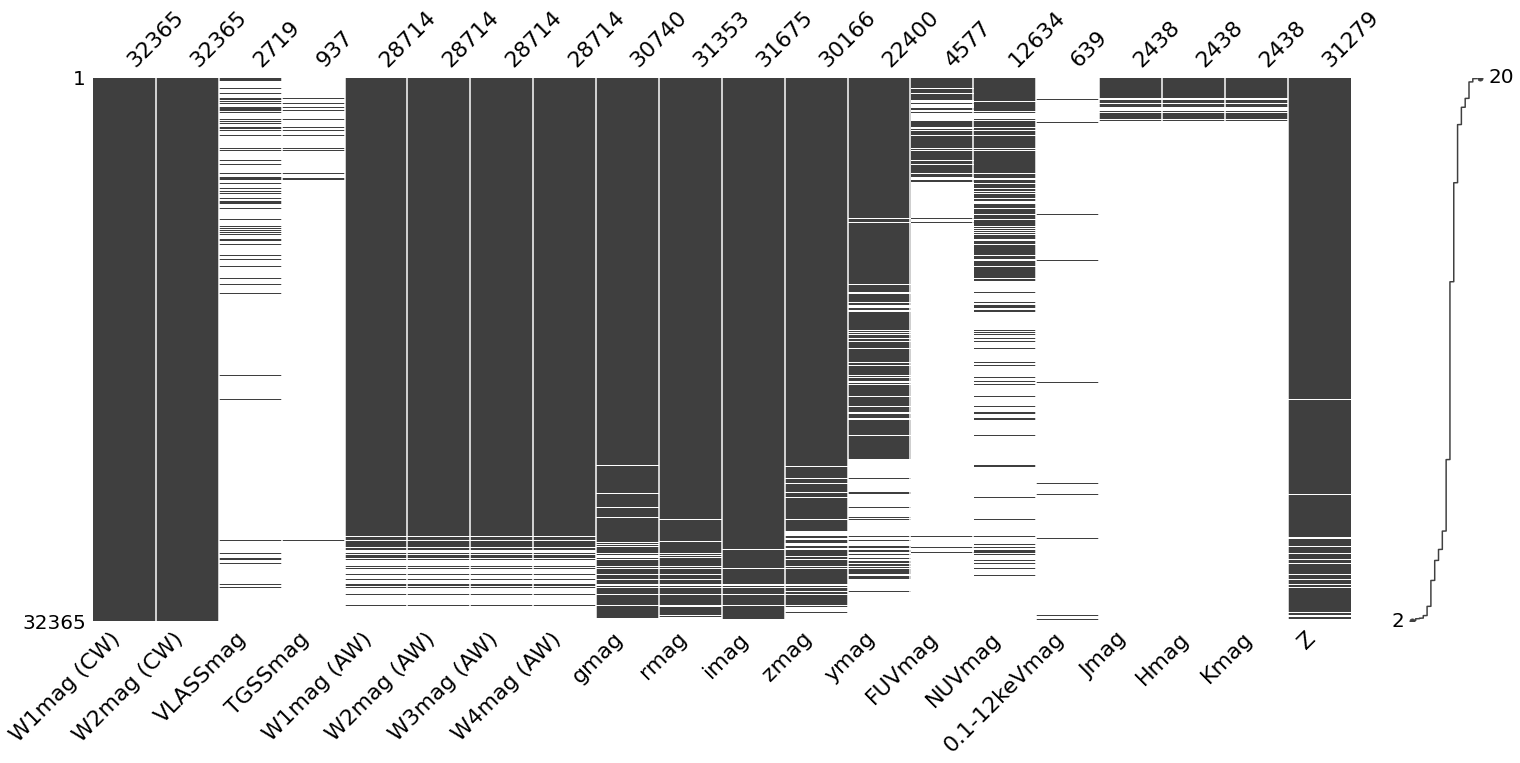}
\caption{Distribution of empty values in HETDEX dataset. Each column shows the data from one feature and dark spaces indicate rows with a valid entry. The~number of valid entries per feature is seen on top of each column. The~dark line in the right side of the plot shows how many measurements each source in the dataset has. For~clarity, sources have been sorted by number of entries, not affecting further~results. \label{fig:missingno_hetdex_matrix}}
\end{figure}

\end{paracol}
\begin{paracol}{2}
\switchcolumn

One important feature to note in Figure~\ref{fig:histograms_hetdex}b is the distribution of radio-detected sources (i.e., sources which show a counterpart on either LOFAR, GMRT, or VLASS). It does not follow the same trends as non-radio AGN, i.e., there is not a peak around $z = 2$ and the number of sources increases towards $z=0$. This behaviour appears from the inclusion of all AGN listed in the MQC. Its documentation\endnote{\url{http://quasars.org/Milliquas-ReadMe.txt} (accessed on 25 October 2021).}
 states that some sources with a strong influence from their host galaxies and QSO candidates are included along with confirmed core-dominated AGN. These objects exhibit, mostly, galaxy-like properties (with low radio emission levels, which might only be detected at redshift values close to $z=0$), shifting their distributions to unusual shapes for AGN. In~addition to that, and~given that only around a $3 \%$ of the SDSS sources in our sample were detected in the radio bands when catalogued \citep{2020ApJS..250....8L}, and~SDSS do not show this misshapen distribution, the~distortion affects mostly radio-detected~sources.

Even though this deviation from the expected distribution might affect, in~some way, part of our results, we will keep the sources that produce it in our calculations. As~mentioned in Section~\ref{sec:intro}, part of the aims of this work is test whether an ML model can deliver reasonable results without discarding, or~modifying, a~large fraction of the intital~dataset.

We also calculated colours for some of the bands. We computed g - r, r - i, i - z, z - y, g - i, W1 - W2, W2 - W3, W3 - W4, J - H, H - K, and~FUV - NUV. In addition,~following the results from \citet{2018A&A...616A..97D, 2021A&A...649A..81N}, who studied different combinations of features and their positive impact on the prediction of redshifts, we have constructed ratios of magnitudes. The~created quantities are r/z, i/y, W1/W3, W1/W4, W2/W4, J/K, and~FUV/K. Finally, we included two indicators, in~the form of a boolean flag, showing whether a source has a measurement on any radio band (LOFAR or TGSS) or on X-ray (Full band in XMM-Newton).

\subsection{Methods}\label{sec:methods}

\subsubsection{Data~Preparation}\label{sec:dataprep}

Redshift values have a logarithmic behaviour when compared to the time passed---and distance travelled---between two values. A~unit difference at low redshift has not the same significance as a unit difference at high redshift (i.e., early epochs). Given that, ultimately, redshifts can be used to determine distances, and~times, from~the observer to a given source, it is useful to make this quantity comparable to linear measurements. Thus, to~overcome this non-linearity and, at~the same time, establish a procedure to contrast predictions and real values, all comparisons will be normalised by the real redshift as follows:
\begin{equation}
\Delta z^{N} = \frac{z_{\mathrm{Predicted}} - z_{{\mathrm{True}}}}{1 + z_{\mathrm{True}}} = \frac{\Delta z}{1 + z_{\mathrm{True}}}.
\end{equation}

Using these two quantities, $\Delta z$ and $\Delta z_{N}$, it is possible to define a set of metrics to assess the quality of the prediction the developed models can achieve. First, we can define the standard deviation between the true, original redshift and the predicted value.
\begin{equation}
\sigma_{\Delta z} = \sqrt{\frac{1}{N} \sum_{i}^{N} \Delta z^{2}}.
\end{equation}

In the same way, the~value $\Delta z^{N}$ can be used instead of $\Delta z$, giving rise to the normalised standard deviation, $\sigma_{\Delta z}^{N}$.

Alternatively, the~redshift deviations can be used directly to create the median absolute deviation (MAD),
\begin{equation}
\sigma_{MAD} = 1.48 \times \mathrm{median} |\Delta z|,
\end{equation}

\noindent or the normalised MAD (NMAD) with~the weighted redshift deviations, $\Delta z^{N}$.

Another quantity used to evaluate the predictions is the fraction of outliers, $\eta$. It represents the number of predictions that are too far away from the true value over the total number of prediction. There are several ways to define this value~\cite{2010A&A...523A..31H, 2010MNRAS.401.1399B, 2021arXiv210401875H, 10.3389/fspas.2021.658229}. We will make use of the interpretation by \citet{2010A&A...523A..31H}, which considers all predictions that fulfil the following condition to be outliers:
\begin{equation}\label{eq:cat_fraction}
\left|\Delta z^{N}\right| > 0.15.
\end{equation}

Using both the standard and normalised differences between redshift values can allow us to analyse the results of our predictions from two points of view: from a purely statistical standpoint (using the standard difference), and~a physically-motivated perspective, with~the use of the normalised redshift difference. Both approaches can be useful to reach a better understanding of the behaviour of the used~models.

For this work, we have analysed our data using the Regression module of the Python package \verb_PyCaret_\endnote{\url{https://pycaret.org} (accessed on 23 October 2021).}
 \cite{PyCaret}. It can create a full pipeline for the use of our dataset and has enough options to change its parameters as~needed.

The first step of data preparation is imputation. A~large fraction of ML models cannot be used with missing data. For~this reason, several methods have been devised to impute missing values (for a review on data imputation, see, for~instance, Reference \cite{ChattopadhyayData}). In~our dataset, several features have a large fraction of empty spaces. A~distribution of empty entries, prepared with the software \verb_missingno_~\cite{aleksey_bilogur_2021_5068743}, can be seen in Figure~\ref{fig:missingno_hetdex_matrix}. It is possible to see that radio and X-ray features have the largest number of empty~values.

We imputed each magnitude with its detection limit and propagated those values for colours and ratios, assuming that empty entries are faint enough to be detected by each instrument. Thereafter, and~within the \verb_PyCaret_ frame, we further removed features based on their influence over the prediction. We applied the Boruta method~\cite{JSSv036i11}, discarding a feature if it behaves better than an aleatory version of itself. The~final list of used features is seen in Table~\ref{tab:feat_imp_redshift}.  The~remaining features are re-scaled to have a mean value of $\mu = 0$ and a standard deviation of $\sigma = 1$ and, afterwards, power-transformed to resemble a Gaussian distribution using the Yeo-Johnson method~\cite{10.1093/biomet/87.4.954}. The~use of re-scaling steps helped our models to improve their results over training with the original features. No further modifications were applied to the data. Thus, no corrections are applied for obscuration, AGN variability, host galaxy morphology, or~other~properties.

For the validation of our ML model, we have set aside a $10 \%$ of the full dataset. From~the remaining $90 \%$,~$70 \%$ was used for training, and~$30 \%$ for~model testing. The~same distribution of sources, which was created randomly, was used throughout the full study. Following the conventions used by \verb_PyCaret_, the~validation sub-set is the only fraction of the data which is not used for the training~stage.

\end{paracol}
\begin{paracol}{2}
\switchcolumn
\vspace{-12pt}


\subsubsection{Model Selection and~Stacking}\label{sec:modelselect}


With the help of \verb_PyCaret_, we run simple realisations of a list of known ML model and selected, as~meta-learner, the~model with the best score ($\sigma_{z}^{N}$, see Section~\ref{sec:dataprep}). After~these tests, we stacked the four models with the following best metrics. Model stacking takes the results (predictions) from several models and adds them as new features for the meta-learner. In~this way, the~meta-learner can use the properties and advantages of the remaining models as a guidance for its own training and improve the prediction results. Furthermore, stacking can help improving the overall scores of the predictions. The~stacked model was trained using 10-fold Cross Validation. The~metrics of the training of the base and meta learners, along with those from the stacked model are presented in Table~\ref{tab:scores_z_stack}.

\begin{specialtable}[H]
\caption{Model Stacking results. Only $\sigma_{NMAD}$ was used to rank the models and select base and meta learners (see Section~\ref{sec:modelselect}). Stacked Train refers to the use of the stacked model in the training set and Stacked Train+Test to the same model in the union of training and test~sets. \label{tab:scores_z_stack}}
\resizebox{1\columnwidth}{!}{
\begin{tabular}{cccccccc}
\toprule
    & \textbf{Random} & \textbf{Extra} & \textbf{CatBoost} & \textbf{LightGBM} & \textbf{XGBoost} & \textbf{Stacked} & \textbf{Stacked}\\
    & \textbf{Forest} & \textbf{Trees} & \textbf{} & \textbf{} & \textbf{} & \textbf{Train} & \textbf{Train + Test}\\
\midrule
\textbf{$\sigma_{NMAD}$}  & $0.1040$ & $0.1079$ & $0.1225$ & $0.1251$ & $0.1295$ & $0.0971$ & $0.1000$\\
\textbf{$\sigma_{z}^{N}$}  & $0.4639$ & $0.4608$ & $0.4587$ & $0.4656$ & $0.4771$ & $0.4495$ & $0.4445$\\
\bottomrule
\end{tabular}}
\end{specialtable}

\section{Results}\label{sec:results}
\unskip

\subsection{Redshift~Prediction}\label{sec:results_z}

For the model stacking, we have chosen, as~base models, \verb_Extra Trees_ (Extremely Randomised Trees, \cite{Geurts2006}), \verb_CatBoost_~\cite{DBLP:journals/corr/DorogushGGKPV17, DBLP:journals/corr/abs-1810-11363}, \verb_LightGBM_~\cite{NIPS2017_6449f44a}, and~\verb_XGBoost_~\cite{Chen:2016:XST:2939672.2939785}. A~\verb_Random Forest_ regression model~\cite{Breiman2001} was used as meta-learner. From~the $10$-fold Cross Validation training, we have obtained a value of $\sigma_{NMAD} = 0.0971 \pm 0.0027$ (see Table~\ref{tab:scores_z_stack}, where the~uncertainty value corresponds to the standard deviation of the Cross Validation instances). This is in the order of a one hundredth of a scaled redshift unit, improving upon the results from the individual models. In addition, when including the test set in the training of the model, the~normalised standard deviation is $\sigma_{NMAD} = 0.1000$.

Figure~\ref{fig:prediction_z_plots}a shows the prediction values for the validation set. The~density of the plotted points, with~higher values shown as a darker colours, shows that a large fraction of the predictions are close to the $y~{=}~x$ line. Additionally, the~outlier fraction (Equation~\eqref{eq:cat_fraction}) for the HETDEX Spring Field validation sample is $\eta = 21.87\%$. The~results of the prediction over the test and validation sets are summarised in Table~\ref{tab:scores_z_full}.

\end{paracol}
\begin{figure}[H]
\widefigure
\begin{subfigure}[b]{0.49\columnwidth}
\includegraphics[width=\textwidth]{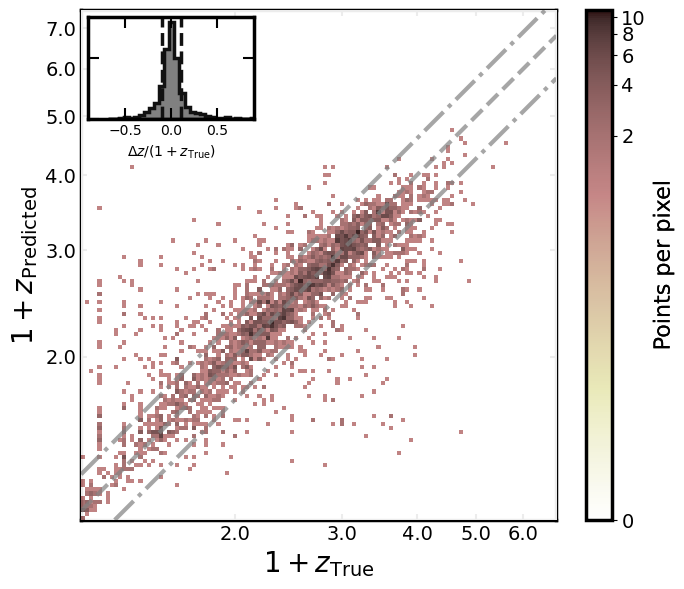}
\caption{HETDEX Spring Field validation~set.}
\label{fig:prediction_z_validation}
\end{subfigure}
\hfill
\begin{subfigure}[b]{0.49\columnwidth}
\includegraphics[width=\textwidth]{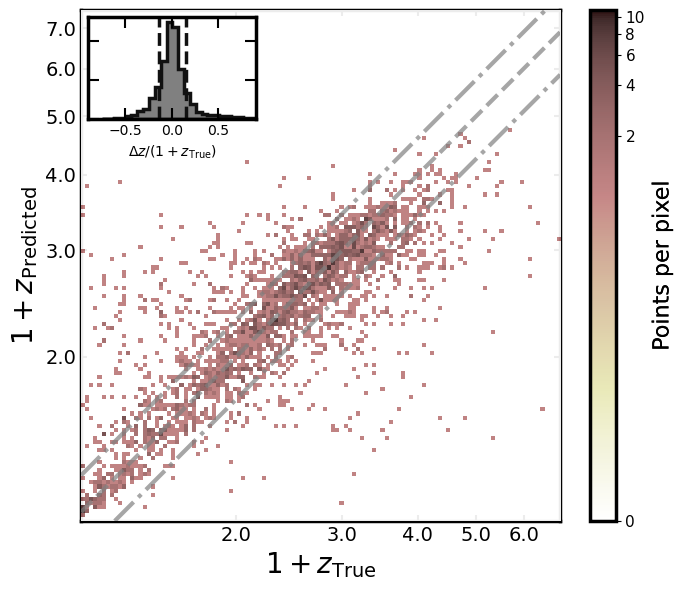}
\caption{Stripe 82 Field~sample.}
\label{fig:prediction_z_s82}
\end{subfigure}
\vspace{6pt}

\caption{Distribution of predicted redshifts as a function of the original redshift values from the validation sample and Stripe 82 Field. Each square represents the number of sources as colour-coded in the colour bar. Diagonal, dashed line represents the $x = y$ relation and the dotted dashed lines show the zone of outliers. The~panel in the upper-left side of each figure shows the distribution of $\Delta z^{N}$ values from the~prediction.}\label{fig:prediction_z_plots}
\end{figure}
\begin{paracol}{2}
\switchcolumn
\vspace{-12pt}

\begin{specialtable}[H]
\caption{Results from the application of the model to the Test and Validation sets, to~the full Stripe 82 sample, and~to the cross-match between our sample and the X-ray sources from \mbox{\citet{2017ApJ...850...66A}} (see Section~\ref{sec:prev_results}).\label{tab:scores_z_full}}

\setlength{\cellWidtha}{\columnwidth/5-2\tabcolsep+.0in}
\setlength{\cellWidthb}{\columnwidth/5-2\tabcolsep-0in}
\setlength{\cellWidthc}{\columnwidth/5-2\tabcolsep-0.0in}
\setlength{\cellWidthd}{\columnwidth/5-2\tabcolsep-0.0in}
\setlength{\cellWidthe}{\columnwidth/5-2\tabcolsep-0.0in}
\scalebox{1}[1]{\begin{tabularx}{\columnwidth}{>{\PreserveBackslash\centering}m{\cellWidtha}>{\PreserveBackslash\centering}m{\cellWidthb}>{\PreserveBackslash\centering}m{\cellWidthc}>{\PreserveBackslash\centering}m{\cellWidthd}>{\PreserveBackslash\centering}m{\cellWidthe}}
\toprule
    & \textbf{HETDEX} & \textbf{HETDEX} & \textbf{Stripe 82} & \textbf{Stripe 82}\\
    & \textbf{Test Set} & \textbf{Validation Set} & \textbf{Test Set} & \textbf{Ananna+17}\\
\midrule
\textbf{$\sigma_{MAD}$}  & $0.1392$ & $0.2118$ & $0.2854$ & $0.2287$\\
\textbf{$\sigma_{NMAD}$}  & $0.0594$ & $0.0906$ & $0.1197$ & $0.1122$\\
\textbf{$\sigma_{z}$}  & $0.2756$ & $0.4287$ & $0.5528$ & $0.3630$\\
\textbf{$\sigma_{z}^{N}$}  & $0.1162$ & $0.1986$ & $0.2501$ & $0.1834$\\
\textbf{$\eta$}  & $0.1158$ & $0.2187$ & $0.2972$ & $0.2429$\\
\bottomrule
\end{tabularx}}
\end{specialtable}

\subsection{Prediction in Stripe 82~Field}\label{sec:S82_preds}

To avoid possible biases derived from predicting on the same type of data as that used in training, and~to test the prediction capabilities of our model, we applied it in data from a different area of the sky. In~this case, we selected the SDSS Stripe 82 Field. We gathered the same data as described in Section~\ref{sec:data}. The~main difference is that this field is not covered by the LoTSS-DR1 Survey. Thus, the~selected area is defined by the coverage of the VLA SDSS Stripe 82 Survey~\cite{2011AJ....142....3H}. This is to mimic the use, as~with the HETDEX Spring Field, of~an area covered by a radio survey. The~VLA-Stripe 82 Survey covers an area over $92$ $\mathrm{deg}^{2}$ with a median rms noise of $52$ $\upmu$J/beam and an angular resolution of $1.8 \arcsec$. We have selected this field because of the high-quality measurements it hosts, and~thorough studies on AGN over its area. The~sample we have produced has 369,093 detected sources and 2941 of them have been labelled as AGN by the MQC. Additionally, $111$ sources have been defined as QSO~candidates.


In Figure~\ref{fig:prediction_z_plots}b, the~results of the redshift predictions, along with the original values, for~Stripe 82 are presented. Results from Stripe 82, shown in Table~\ref{tab:scores_z_full}, resemble those of the HETDEX Spring Field, hinting the possibility of, as~long as the needed wavebands are available, using the trained models in areas of the sky which are not related to the training~sample.

In addition, and~even though all metric results are better in the initial HETDEX Spring sample, differences with Stripe 82 are on the range of 7--8\%. These deviations are small enough to be caused by statistical variations among both fields. In~the case of the outlier fraction, it is around $30 \%$, and~$8$ percentage points higher than with the primary~sample.

\section{Discussion}\label{sec:discussion}
\unskip

\subsection{Previous~Results}\label{sec:prev_results}

As a way to assess our results, it is possible to compare them to previous redshift determinations. This is the case of \citet{2017ApJ...850...66A}. They used multiwavelength data from 5961 X-ray-detected AGN in the Stripe 82 Field with $z \leq 3.0$ and, from~fitting SED models, they computed photometric redshifts. From~their Table~7, a~value of $\sigma_{NMAD} = 0.0602$ is quoted for their full sample, which is in line with our prediction in the Stripe 82 Field ($\sigma_{NMAD} = 0.1197$). In addition,~an outlier fraction of $13.69 \%$ is achieved, less than half of what is obtained using our stacked model in the same area. It is possible to select, from~our sample, the~sources with a counterpart in the \citet{2017ApJ...850...66A} sample and apply our model to them. Using a matching radius of $2 \arcsec$, $221$ sources are selected, reaching values of $\sigma_{NMAD} = 0.1122$ and $\eta = 0.2429$. If~we do the same exercise, selecting the results from the SED-fitting redshift determination, their values are $\sigma_{NMAD} = 0.0648$ and $\eta = 0.2048$. Full results for this sub-sample are shown in Table~\ref{tab:quoted_z_scores}.

To contrast our results with previous ML implementations, we can take the work from \citet{2021MNRAS.503.2639C}, who compared the results of applying deep learning, decision trees, and~k-nearest neighbours regression to predict redshift values for $100,000$ SDSS DR12 QSO with accurate spectroscopic redshifts. Results are presented in Table~\ref{tab:quoted_z_scores}.

\begin{specialtable}[H]
\caption{Results from previous works. First column: full X-ray selected sample quoted from \mbox{\citet{2017ApJ...850...66A}}. Second column: selection of sources from \citet{2017ApJ...850...66A} that have a match in our sample. Following columns: result of application of k-Nearest Neigbours (KN), Decision Tree Regression (DT), and~Deep Learning (DL) models as quoted from \citet{2021MNRAS.503.2639C}.\label{tab:quoted_z_scores}}

\setlength{\cellWidtha}{\columnwidth/6-2\tabcolsep-0.3in}
\setlength{\cellWidthb}{\columnwidth/6-2\tabcolsep+.1in}
\setlength{\cellWidthc}{\columnwidth/6-2\tabcolsep+.2in}
\setlength{\cellWidthd}{\columnwidth/6-2\tabcolsep-0.0in}
\setlength{\cellWidthe}{\columnwidth/6-2\tabcolsep-0.0in}
\setlength{\cellWidthf}{\columnwidth/6-2\tabcolsep-0.0in}
\scalebox{1}[1]{\begin{tabularx}{\columnwidth}{>{\PreserveBackslash\centering}m{\cellWidtha}>{\PreserveBackslash\centering}m{\cellWidthb}>{\PreserveBackslash\centering}m{\cellWidthc}>{\PreserveBackslash\centering}m{\cellWidthd}>{\PreserveBackslash\centering}m{\cellWidthe}>{\PreserveBackslash\centering}m{\cellWidthf}}
\toprule
    & \textbf{Stripe 82 Full} & \textbf{Stripe 82 Match} & \textbf{SDSS KN}  & \textbf{SDSS DT}  & \textbf{SDSS DL}\\
    & \textbf{Ananna+17} & \textbf{Ananna+17} & \textbf{Curran+2021} & \textbf{Curran+2021} & \textbf{Curran+2021}\\
\midrule
\textbf{$\sigma_{MAD}$}     & $\cdots$ & $0.1336$ & $0.2360$ & $0.1290$ & $0.0920$ \\
\textbf{$\sigma_{NMAD}$}    & $0.0602$ & $0.0648$ & $0.0500$ & $0.0580$ & $0.0420$ \\
\textbf{$\sigma_{z}$}       & $\cdots$ & $0.5435$ & $0.2360$ & $0.3330$ & $0.2350$ \\
\textbf{$\sigma_{z}^{N}$}   & $\cdots$ & $0.2766$ & $0.1210$ & $0.1600$ & $0.1100$ \\
\textbf{$\eta$}             & $0.1369$ & $0.2048$ & $\cdots$ & $\cdots$ & $\cdots$ \\
\bottomrule
\end{tabularx}}
\end{specialtable}

When comparing our results (Table~\ref{tab:scores_z_full}) with the outputs from \citet{2021MNRAS.503.2639C}, we note that the metrics for our Validation set are 20--40\% higher and those from the Stripe 82 Field, 40{--}60\% higher than theirs. This is a consequence of our decision of not cleaning our training set, mimicking the conditions a large dataset might present. They, in~contrast, have trained their models with sources that have full coverage on the bands they selected, avoiding the use of imputation. Moreover, since they have used large SDSS sample, the~properties of QSO among them are more homogeneous than that of the present work, leading to improved prediction~results.

Comparison with previous works, using traditional template-based and ML photometric redshift determination methods, highlights the prospective scenarios to apply our model. Rather than selecting a very small area with the right conditions, we can use the model here presented on large regions with incomplete coverage, rising the likelihood of obtaining objects with specific resdhift~values.

\subsection{Feature~Importances}\label{sec:feat_importances}


Feature importances from our model are listed in Table~\ref{tab:feat_imp_redshift}. The~values are provided by the model itself, and they have been calculated as the mean decrease on impurity for the ensemble of trees. We can see that the features with the highest importances are those coming from the CatWISE catalogue. After~them, quantities are 
derived from Pan-STARRS. In addition, finally, those obtained from AllWISE, and~GALEX observations 
suggest a very low impact in the model training and the predictions derived from~it.

\begin{specialtable}[H]
\caption{Features used by our redshift prediction model and their~importances.\label{tab:feat_imp_redshift}}
\resizebox{1\columnwidth}{!}{
\begin{tabular}{lrclrclr}
\toprule
\textbf{Feature}	& \textbf{Importance} &	& \textbf{Feature} & \textbf{Importance} &	& \textbf{Feature} & \textbf{Importance}\\
\midrule
W1 - W2 (CW)    &   87.381  &   &	z - y           &	37.084 &    &   FUV - NUV   &   11.338\\
W1 (CW)         &   82.759  &   &	W1/W3 (AW)      &	33.207 &    &   FUV/K       &    8.886\\
g - i           &   70.617  &   &	i/y             &	33.081 &    &   FUV         &    7.202\\
g               &   55.787  &   &	W2/W4 (AW)      &	29.196 &    &   K           &    5.484\\
W2 - W3 (AW)    &   53.919  &   &	i - z           &	28.647 &    &   J - H       &    2.817\\
r/z             &   52.251  &   &	W4 (AW)         &	26.392 &    &   J/K         &    2.803\\
y               &   49.234  &   &	W3 - W4 (AW)    &	24.898 &    &   H - K       &    2.771\\
r - i           &   46.451  &   &	NUV             &	23.296 &    &               &     \\
\bottomrule
\end{tabular}}
\end{specialtable}

Entries from CatWISE have the largest amount of relevant, non-repetitive information from all features. Despite the different nature of the used features, i.e., magnitudes, colours, ratios, there is no clear preference of one kind over the others. The~main factor to have high importance is the fraction of sources with a measurement in the studied feature. This distribution also reinforces the results from Reference~\cite{2018A&A...616A..97D}, who established that is possible to use combinations of magnitudes other than colours and train, successfully, ML~models.

Table~\ref{tab:feat_imp_redshift} also gives information on the features that can be discarded from the model training without having a high influence on the predicted values (features with data from 2MASS and GALEX). Finally, it is important to stress that, in~this work, we have not discarded data based upon the feature~importances.

\subsection{Shapley~Explanations}\label{sec:shap_explanations}


Shapley values were obtained using the Tree-based module of the Python package \verb_SHAP_\endnote{\url{https://github.com/slundberg/shap} (accessed on 18 October 2021).}
 \cite{NIPS2017_7062, lundberg2020local2global}. In~Figure~\ref{fig:shap_vals_z_hetdex}, features are sorted by decreasing median Shapley~values.

The quantity with the highest Shapley value is related to the base observations. However,~from the distribution of values in the horizontal axis for the W1 magnitude, it is possible to see that its large dispersion implies that its influence on predictions can drive the final redshift either to low or high values. This is in contrast with, for~instance, the~g - i colour. Its Shapley values might be close to zero, indicating that it does not have impact on the redshfit prediction. The~values can be higher than zero, as~well, driving the predictions to high redshift~values.

\begin{figure}[H]
\includegraphics[width=0.75\columnwidth]{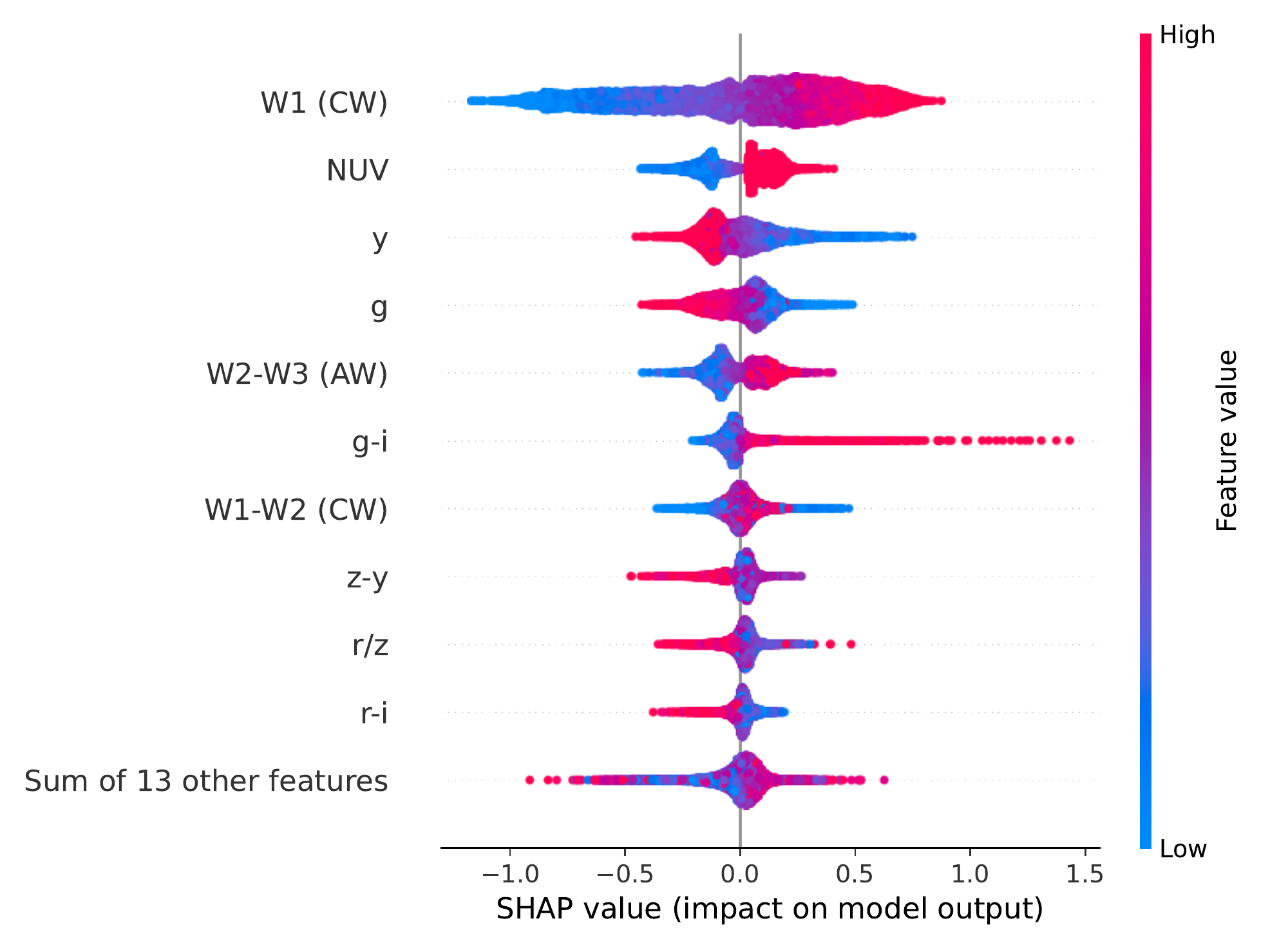}
\caption{Shapley values for the ten features with the highest median Shapley numbers in our redshift prediction model. Each row corresponds to one feature. Colour map indicates the value of the feature for each source. Features are sorted by median Shapley value. Last row shows the sum of the 13~remaining features used for the prediction. Feature values of points to the left of vertical, grey line have a positive impact on the model output, i.e., redshift will tend to be higher. Points close to the vertical line show a limited impact on the redshift~prediction. \label{fig:shap_vals_z_hetdex}}
\end{figure}

Most of the remaining features show Shapley values clustered around $0.0$, and~a small sub-sample deviates from this and has a noteworthy influence on~predictions.

The feature with the second highest median Shapley values is the NUV magnitude from GALEX. From~Figure~\ref{fig:missingno_hetdex_matrix}, it is possible to see that this feature exhibits a very high fraction of empty entries. That implies that most of sources have an imputed NUV magnitude. This distribution is present in Figure~\ref{fig:shap_vals_z_hetdex}. Therefore, all imputed magnitudes make the redshift prediction go up, and~all measured magnitudes make it go down. Although~this behaviour might seem anomalous, it has its roots on the fact that very few high-redshift sources are detectable by~GALEX.

Being able to retrieve these interpretations is one of the advantages of using Shapley values from a prediction model. It is possible to understand whether certain range of values of a feature can make a prediction go up or down. This differs from feature importances, which allow an average view of the impact of a feature over the complete trained~model.

Despite their differences, feature importances and Shapley values can help understand the impact that measurements in different wavelengths can have over the understanding and prediction of redshift values of AGN. In~particular, and~given the relevance and high-quality observations that future radio surveys and observatories will deliver, adding direct measurements (e.g., Reference \cite{2020MNRAS.499.3660T}) or features derived from them might be highly beneficial when focusing the search on high-redshift objects. The~latter might be the case with already-known quantities, such as radio loudness or radio spectral indices. These properties can provide indications on the radio emission~\cite{2019A&A...630A..83Z} and its relation with other \mbox{wavelengths~\cite{2008MNRAS.390..847L, 2019MNRAS.482.5513L}}.

\section{Conclusions}\label{sec:conclusions}

In this work, we trained several Machine Learning models to predict, from~a sample of infrared-detected AGN---and their multiwavelength counterparts---their redshift~value.

Sources were obtained from CatWISE2020 catalogue and counterpart measurements were obtained from AllWISE, Pan-STARRS, LOFAR, GMRT, VLASS, GALEX, 2MASS, and~XMM-NEWTON observations and surveys. All of the sources are located at the HETDEX Spring~Field.

Using of the \verb_PyCaret_ Python package as a framework, we stacked four different models with a meta-learner. The~application of model to the validation set lead a median redshift error on the prediction of $\sigma_{z}^{N} = 0.1986$ and an outlier fraction of $\eta = 21.87 \%$. This goes in line with previous results, taking into account that no major cleaning procedure was performed into the~dataset.

To further test the power of our model, we applied it to a separate catalogue of AGN located in the Stripe 82 Field, and the median redshift error was $\sigma_{z}^{N} = 0.2501$ and an outlier fraction of $\eta = 29.72 \%$.

To understand the influence of the different features included in the model, Shapley values were calculated for the training sub-set. The~features from WISE and from Pan-STARRS show the highest median Shapley values, mirroring the fact that these features have the lowest number of imputed~entries.

The results presented in this work stress the benefits of using ML as an initial approach to derive redshift predictions for AGN. Using a fraction of the time a template-based photometric redshift determination tool might take, ML can give redshift predictions with a high confidence level which can lead to further studies of selected sources. This advantage might become critical to the use of current and future large-area surveys---with radio surveys being a major example, which need to extract information from several millions of sources within an appropriate amount of~time.

Even though some of the results obtained in this work do not show a considerable improvement from previous studies, it is relevant to emphasise that our work was aimed to extract predictions using datasets without large amounts of preparation, i.e., feature engineering. This implies that it is possible to use a very heterogeneous group of datasets (with different sensitivities, resolutions, etc.) and obtain useful predictions from them without the need of cleaning and reducing the number of used sources in each~catalogue.

Our model can be further improved using future surveys which will cover large areas with very deep observations. One such survey is Data Release 2 of the LoTSS survey, which will be released in the near future. It will cover 5720 $\mathrm{deg}^{2}$ in the northern sky with similar sensitivities as DR1~\cite{2021MNRAS.505L..36M}. If~assuming the same AGN density as in LoTSS DR1 (see Section~\ref{sec:data}, with~32,365 AGN in $424 \, \mathrm{deg}^{2}$), DR2 is expected to deliver 436,622 AGN from its area. This will allow us training a redshift prediction model with a number of sources one order of magnitude larger, improving its accuracy dramatically by capturing the properties from a larger parameter space. This improvement can be also analysed in terms of cosmic variance. Following the results by Reference~\cite{2010MNRAS.407.2131D}, DR1 from the LoTSS survey will be subject to a cosmic variance between $10$ and $20\%$. In addition, extrapolating the curve from their Figure~6, DR2 will make this value go below $10\%$. Only from this improvement, we might expect to achieve a better training for a prediction model. AGN might present variability on their observations with different timescales~\cite{2004A&A...421..913W, 2009ApJ...690.1250S}, which might impact the observed properties of the used datasets. These variations can increase the fraction of outliers in different ranges~\cite{2012A&A...542A..20M}.

Additional sources of improvement in the results are related to the treatment of the missing values in our catalogue. Devising more advanced imputation methods, which can take into account the distribution of measured values in one feature and their relation to the rest of features, might refine our results. Related to this, some features have a low fraction of measured values, adding little information to the models. Discarding these features also might reduce the fraction of outliers. Apart from the data treatment, further improvements might be achieved if the intrinsic time variability of AGN is taken into~account.

The used model might arrive to better results creating several instances of data sub-sets. Using different combinations of sources for training, test, and~validation might have an impact on how the model arrives to separate~predictions.

With all these advantages, the~model described in this article can be used as part of a full pipeline which might be able to predict the presence of AGN in a large-area field. In addition, for~the predicted AGN, we 
predict their redshift values, among other properties, e.g., radio detectability. This might allow the creation of catalogues with high-redshift Radio Galaxies from datasets covering large~areas.



\vspace{6pt}



\authorcontributions{Conceptualisation, I.M., J.A., and~R.C.; methodology, R.C., P.C., and~A.H.; software, R.C. and~A.H.; validation, I.M, J.A., S.A., and~R.C.; formal analysis, R.C.; investigation, R.C.; resources, J.A. and I.M.; data curation, R.C.; writing---original draft preparation, R.C. and I.M.; writing---review and editing, R.C., J.A., I.M., S.A., D.B., P.C., and~A.H.; visualisation, R.C.; supervision, J.A., I.M.; project administration, I.M. and~J.A. All authors have read and agreed to the published version of the~manuscript.}

\funding{This work was supported by Fundação para a Ciência e a Tecnologia (FCT) through the research grants PTDC/FIS-AST/29245/2017, UID/FIS/04434/2019, UIDB/04434/2020, and UIDP/04434/2020. R.C. acknowledges support from the Fundação para a Ciência e a Tecnologia (FCT) through the Fellowship PD/BD/150455/2019 (PhD:SPACE Doctoral Network PD/00040/2012) and POCH/FSE (EC). A.H. acknowledges support from contract DL 57/2016/CP1364/CT0002 and an FCT-CAPES funded Transnational Cooperation project ``Strategic Partnership in Astrophysics Portugal-Brazil''.}

\institutionalreview{Not applicable.}

\informedconsent{Not applicable.}

\dataavailability{Not applicable}

\acknowledgments{The authors would like to thank the referees for discussion and suggestions leading to the improvement of this work. This publication makes use of data products from the Wide-field Infrared Survey Explorer, which is a joint project of the University of California, Los Angeles, and~the Jet Propulsion Laboratory/California Institute of Technology, funded by the National Aeronautics and Space Administration.
LOFAR data products were provided by the LOFAR Surveys Key Science project (LSKSP\endnote{\url{https://lofar-surveys.org/} (accessed on 3 August 2021).}
) and were derived from observations with the International LOFAR Telescope (ILT). LOFAR~\cite{2013A&A...556A...2V} is the Low Frequency Array designed and constructed by ASTRON. It has observing, data processing, and~data storage facilities in several countries, which are owned by various parties (each with their own funding sources), and~which are collectively operated by the ILT foundation under a joint scientific policy. The~efforts of the LSKSP have benefited from funding from the European Research Council, NOVA, NWO, CNRS-INSU, the~SURF Co-operative, the~UK Science and Technology Funding Council, and the Jülich Supercomputing Centre.
The Pan-STARRS1 Surveys (PS1) and the PS1 public science archive have been made possible through contributions by the Institute for Astronomy, the~University of Hawaii, the~Pan-STARRS Project Office, the~Max-Planck Society and its participating institutes, the~Max Planck Institute for Astronomy, Heidelberg and the Max Planck Institute for Extraterrestrial Physics, Garching, The Johns Hopkins University, Durham University, the~University of Edinburgh, the~Queen's University Belfast, the~Harvard-Smithsonian Center for Astrophysics, the~Las Cumbres Observatory Global Telescope Network Incorporated, the~National Central University of Taiwan, the~Space Telescope Science Institute, the~National Aeronautics and Space Administration under Grant No. NNX08AR22G issued through the Planetary Science Division of the NASA Science Mission Directorate, the~National Science Foundation Grant No. AST-1238877, the~University of Maryland, E\"{o}tv\"{o}s Lor\'{a}nd University (ELTE), the~Los Alamos National Laboratory, and~the Gordon and Betty Moore Foundation.
This research has made use of data obtained from the 4XMM XMM-Newton serendipitous source catalogue compiled by the 10 institutes of the XMM-Newton Survey Science Centre selected by ESA.
This publication makes use of data products from the Two Micron All Sky Survey, which is a joint project of the University of Massachusetts and the Infrared Processing and Analysis Center/California Institute of Technology, funded by the National Aeronautics and Space Administration and the National Science Foundation.
This research has made use of the VizieR catalogue access tool, CDS, Strasbourg, France (DOI : 10.26093/cds/vizier). The~original description of the VizieR service was published in Reference~\cite{vizier}.
This research made use of Astropy\endnote{\url{https://www.astropy.org} (accessed on 23 July 2021).}
, a~community-developed core Python package for Astronomy~\cite{astropy:2013, astropy:2018} and TOPCAT\endnote{\url{http://www.star.bris.ac.uk/~mbt/topcat/} (accessed on 29 July 2021).}
 \cite{2005ASPC..347...29T}.
This research has made use of NASA’s Astrophysics Data System.}

\conflictsofinterest{The authors declare no conflict of~interest.}


\newpage
\abbreviations{Abbreviations}{The following abbreviations are used in this manuscript:\\

\noindent
\begin{tabular}{@{}ll}
AGN & Active Galactic Nuclei\\
QSO & Quasi Stellar Object\\
ML & Machine Learning\\
RG & Radio Galaxy\\
EoR & Epoch of Reionisation\\
CW & CatWISE2020 Catalogue\\
AW & AllWISE Catalogue
\end{tabular}}

\appendixtitles{no} 



\end{paracol}
\begin{adjustwidth}{0.0cm}{0cm}
\printendnotes[custom]
\end{adjustwidth}

\reftitle{References}

%



\begin{thebibliography}{999}

\bibitem[{Padovani} {et~al.}(2017){Padovani}, {Alexander}, {Assef}, {De
Marco}, {Giommi}, {Hickox}, {Richards}, {Smol{\v{c}}i{\'c}},
{Hatziminaoglou}, {Mainieri}, and {Salvato}]{2017A&ARv..25....2P}
{Padovani}, P.; {Alexander}, D.M.; {Assef}, R.J.; {De Marco}, B.; {Giommi}, P.;
{Hickox}, R.C.; {Richards}, G.T.; {Smol{\v{c}}i{\'c}}, V.; {Hatziminaoglou},
E.; {Mainieri}, V.; et al.
\newblock {Active galactic nuclei: What's in a name?}
\newblock {\em Astron. Astrophys. Rev.} {\bf 2017}, {\em 25},~2, doi:10.1007/s00159-017-0102-9.

\bibitem[{Heckman} and {Best}(2014)]{2014ARA&A..52..589H}
{Heckman}, T.M.; {Best}, P.N.
\newblock {The Coevolution of Galaxies and Supermassive Black Holes: Insights
from Surveys of the Contemporary Universe}.
\newblock {\em Annu. Rev. Astron. Astrophys.} {\bf 2014}, {\em 52},~589--660,
 doi:10.1146/annurev-astro-081913-035722.

\bibitem[{McGreer} {et~al.}(2006){McGreer}, {Becker}, {Helfand}, and
{White}]{2006ApJ...652..157M}
{McGreer}, I.D.; {Becker}, R.H.; {Helfand}, D.J.; {White}, R.L.
\newblock {Discovery of a z = 6.1 Radio-Loud Quasar in the NOAO Deep Wide Field
Survey}.
\newblock {\em Astrophys. J.} {\bf 2006}, {\em 652},~157--162,
 doi:10.1086/507767.

\bibitem[{Ku{\'z}micz} and {Jamrozy}(2021)]{2021ApJS..253...25K}
{Ku{\'z}micz}, A.; {Jamrozy}, M.
\newblock {Giant Radio Quasars: Sample and Basic Properties}.
\newblock {\em Astrophys. J.} {\bf 2021}, {\em 253},~25,
 doi:10.3847/1538-4365/abd483.

\bibitem[{Delhaize} {et~al.}(2021){Delhaize}, {Heywood}, {Prescott},
{Jarvis}, {Delvecchio}, {Whittam}, {White}, {Hardcastle}, {Hale}, {Afonso},
{Ao}, {Brienza}, {Br{\"u}ggen}, {Collier}, {Daddi}, {Glowacki}, {Maddox},
{Morabito}, {Prandoni}, {Randriamanakoto}, {Sekhar}, {An}, {Adams}, {Blyth},
{Bowler}, {Leeuw}, {Marchetti}, {Randriamampandry}, {Thorat}, {Seymour},
{Smirnov}, {Taylor}, {Tasse}, and {Vaccari}]{2021MNRAS.501.3833D}
{Delhaize}, J.; {Heywood}, I.; {Prescott}, M.; {Jarvis}, M.J.; {Delvecchio},
I.; {Whittam}, I.H.; {White}, S.V.; {Hardcastle}, M.J.; {Hale}, C.L.;
{Afonso}, J.; et al.
\newblock {MIGHTEE: Are giant radio galaxies more common than we thought?}
\newblock {\em Mon. Not. R. Astron. Soc.} {\bf 2021}, {\em 501},~3833--3845,
 doi:10.1093/mnras/staa3837.

\bibitem[{Lal}(2021)]{2021ApJ...915..126L}
{Lal}, D.V.
\newblock {The Discovery of a Remnant Radio Galaxy in A2065 Using GMRT}.
\newblock {\em Astrophys. J.} {\bf 2021}, {\em 915},~126, doi:10.3847/1538-4357/ac042d.

\bibitem[{Amarantidis} {et~al.}(2019){Amarantidis}, {Afonso}, {Messias},
{Henriques}, {Griffin}, {Lacey}, {Lagos}, {Gonzalez-Perez}, {Dubois},
{Volonteri}, {Matute}, {Pappalardo}, {Qin}, {Chary}, and
{Norris}]{2019MNRAS.485.2694A}
{Amarantidis}, S.; {Afonso}, J.; {Messias}, H.; {Henriques}, B.; {Griffin}, A.;
{Lacey}, C.; {Lagos}, C.d.P.; {Gonzalez-Perez}, V.; {Dubois}, Y.;
{Volonteri}, M.; et al.
\newblock {The first supermassive black holes: Indications from models for
future observations}.
\newblock {\em Mon. Not. R. Astron. Soc.} {\bf 2019}, {\em 485},~2694--2709,
 doi:10.1093/mnras/stz551.

\bibitem[{Thomas} {et~al.}(2021){Thomas}, {Dav{\'e}}, {Jarvis}, and
{Angl{\'e}s-Alc{\'a}zar}]{2021MNRAS.503.3492T}
{Thomas}, N.; {Dav{\'e}}, R.; {Jarvis}, M.J.; {Angl{\'e}s-Alc{\'a}zar}, D.
\newblock {The radio galaxy population in the SIMBA simulations}.
\newblock {\em Mon. Not. R. Astron. Soc.} {\bf 2021}, {\em 503},~3492--3509,
 doi:10.1093/mnras/stab654.

\bibitem[{Bonaldi} {et~al.}(2019){Bonaldi}, {Bonato}, {Galluzzi},
{Harrison}, {Massardi}, {Kay}, {De Zotti}, and {Brown}]{2019MNRAS.482....2B}
{Bonaldi}, A.; {Bonato}, M.; {Galluzzi}, V.; {Harrison}, I.; {Massardi}, M.;
{Kay}, S.; {De Zotti}, G.; {Brown}, M.L.
\newblock {The Tiered Radio Extragalactic Continuum Simulation (T-RECS)}.
\newblock {\em Mon. Not. R. Astron. Soc.} {\bf 2019}, {\em 482},~2--19,
 doi:10.1093/mnras/sty2603.

\bibitem[{Prandoni} and {Seymour}(2015)]{2015aska.confE..67P}
{Prandoni}, I.; {Seymour}, N.
\newblock {Revealing the Physics and Evolution of Galaxies and Galaxy Clusters
with SKA Continuum Surveys}.
\newblock In Proceedings of the Advancing Astrophysics with the Square Kilometre Array (AASKA14), Giardini Naxos, Italy, 9--13 June 2014; p.~67.

\bibitem[{Inayoshi} {et~al.}(2020){Inayoshi}, {Visbal}, and
{Haiman}]{2020ARA&A..58...27I}
{Inayoshi}, K.; {Visbal}, E.; {Haiman}, Z.
\newblock {The Assembly of the First Massive Black Holes}.
\newblock {\em Annu. Rev. Astron. Astrophys.} {\bf 2020}, {\em 58},~27--97,
 doi:10.1146/annurev-astro-120419-014455.

\bibitem[{Ross} and {Cross}(2020)]{2020MNRAS.494..789R}
{Ross}, N.P.; {Cross}, N.J.G.
\newblock {The near and mid-infrared photometric properties of known redshift z
{\ensuremath{\geq}} 5 quasars}.
\newblock {\em Mon. Not. R. Astron. Soc.} {\bf 2020}, {\em 494},~789--803,
 doi:10.1093/mnras/staa544.

\bibitem[{Miley} and {De Breuck}(2008)]{2008A&ARv..15...67M}
{Miley}, G.; {De Breuck}, C.
\newblock {Distant radio galaxies and their environments}.
\newblock {\em Astron. Astrophys. Rev.} {\bf 2008}, {\em 15},~67--144,
 doi:10.1007/s00159-007-0008-z.

\bibitem[{Helfand} {et~al.}(2015){Helfand}, {White}, and
{Becker}]{2015ApJ...801...26H}
{Helfand}, D.J.; {White}, R.L.; {Becker}, R.H.
\newblock {The Last of FIRST: The Final Catalog and Source Identifications}.
\newblock {\em Astrophys. J.} {\bf 2015}, {\em 801},~26,
 doi:10.1088/0004-637X/801/1/26.

\bibitem[{Norris} {et~al.}(2011){Norris}, {Hopkins}, {Afonso}, {Brown},
{Condon}, {Dunne}, {Feain}, {Hollow}, {Jarvis}, {Johnston-Hollitt}, {Lenc},
{Middelberg}, {Padovani}, {Prandoni}, {Rudnick}, {Seymour}, {Umana},
{Andernach}, {Alexander}, {Appleton}, {Bacon}, {Banfield}, {Becker}, {Brown},
{Ciliegi}, {Jackson}, {Eales}, {Edge}, {Gaensler}, {Giovannini}, {Hales},
{Hancock}, {Huynh}, {Ibar}, {Ivison}, {Kennicutt}, {Kimball}, {Koekemoer},
{Koribalski}, {L{\'o}pez-S{\'a}nchez}, {Mao}, {Murphy}, {Messias},
{Pimbblet}, {Raccanelli}, {Randall}, {Reiprich}, {Roseboom},
{R{\"o}ttgering}, {Saikia}, {Sharp}, {Slee}, {Smail}, {Thompson}, {Urquhart},
{Wall}, and {Zhao}]{2011PASA...28..215N}
{Norris}, R.P.; {Hopkins}, A.M.; {Afonso}, J.; {Brown}, S.; {Condon}, J.J.;
{Dunne}, L.; {Feain}, I.; {Hollow}, R.; {Jarvis}, M.; {Johnston-Hollitt}, M.;
et~al.
\newblock {EMU: Evolutionary Map of the Universe}.
\newblock {\em Publ. Astron. Soc. Aust.} {\bf 2011}, {\em 28},~215--248,
 doi:10.1071/AS11021.

\bibitem[{Gordon} {et~al.}(2020){Gordon}, {Boyce}, {O'Dea}, {Rudnick},
{Andernach}, {Vantyghem}, {Baum}, {Bui}, and
{Dionyssiou}]{2020RNAAS...4..175G}
{Gordon}, Y.A.; {Boyce}, M.M.; {O'Dea}, C.P.; {Rudnick}, L.; {Andernach}, H.;
{Vantyghem}, A.N.; {Baum}, S.A.; {Bui}, J.P.; {Dionyssiou}, M.
\newblock {A Catalog of Very Large Array Sky Survey Epoch 1 Quick Look
Components, Sources, and Host Identifications}.
\newblock {\em Res. Notes Am. Astron. Soc.} {\bf 2020},
{\em 4},~175, doi:10.3847/2515-5172/abbe23.

\bibitem[{Shimwell} {et~al.}(2019){Shimwell}, {Tasse}, {Hardcastle},
{Mechev}, {Williams}, {Best}, {R{\"o}ttgering}, {Callingham}, {Dijkema}, {de
Gasperin}, {Hoang}, {Hugo}, {Mirmont}, {Oonk}, {Prandoni}, {Rafferty},
{Sabater}, {Smirnov}, {van Weeren}, {White}, {Atemkeng}, {Bester},
{Bonnassieux}, {Br{\"u}ggen}, {Brunetti}, {Chy{\.z}y}, {Cochrane}, {Conway},
{Croston}, {Danezi}, {Duncan}, {Haverkorn}, {Heald}, {Iacobelli}, {Intema},
{Jackson}, {Jamrozy}, {Jarvis}, {Lakhoo}, {Mevius}, {Miley}, {Morabito},
{Morganti}, {Nisbet}, {Orr{\'u}}, {Perkins}, {Pizzo}, {Schrijvers}, {Smith},
{Vermeulen}, {Wise}, {Alegre}, {Bacon}, {van Bemmel}, {Beswick}, {Bonafede},
{Botteon}, {Bourke}, {Brienza}, {Calistro Rivera}, {Cassano}, {Clarke},
{Conselice}, {Dettmar}, {Drabent}, {Dumba}, {Emig}, {En{\ss}lin}, {Ferrari},
{Garrett}, {G{\'e}nova-Santos}, {Goyal}, {G{\"u}rkan}, {Hale}, {Harwood},
{Heesen}, {Hoeft}, {Horellou}, {Jackson}, {Kokotanekov}, {Kondapally},
{Kunert-Bajraszewska}, {Mahatma}, {Mahony}, {Mandal}, {McKean}, {Merloni},
{Mingo}, {Miskolczi}, {Mooney}, {Nikiel-Wroczy{\'n}ski}, {O'Sullivan},
{Quinn}, {Reich}, {Roskowi{\'n}ski}, {Rowlinson}, {Savini}, {Saxena},
{Schwarz}, {Shulevski}, {Sridhar}, {Stacey}, {Urquhart}, {van der Wiel},
{Varenius}, {Webster}, and {Wilber}]{2019A&A...622A...1S}
{Shimwell}, T.W.; {Tasse}, C.; {Hardcastle}, M.J.; {Mechev}, A.P.; {Williams},
W.L.; {Best}, P.N.; {R{\"o}ttgering}, H.J.A.; {Callingham}, J.R.; {Dijkema},
T.J.; {de Gasperin}, F.; et~al.
\newblock {The LOFAR Two-metre Sky Survey. II. First data release}.
\newblock {\em Astron. Astrophys.} {\bf 2019}, {\em 622},~A1,
 doi:10.1051/0004-6361/201833559.

\bibitem[{Singh} {et~al.}(2014){Singh}, {Beelen}, {Wadadekar}, {Sirothia},
{Ishwara-Chandra}, {Basu}, {Omont}, {McAlpine}, {Ivison}, {Oliver}, {Farrah},
and {Lacy}]{2014A&A...569A..52S}
{Singh}, V.; {Beelen}, A.; {Wadadekar}, Y.; {Sirothia}, S.; {Ishwara-Chandra},
C.H.; {Basu}, A.; {Omont}, A.; {McAlpine}, K.; {Ivison}, R.J.; {Oliver}, S.;
et~al.
\newblock {Multiwavelength characterization of faint ultra steep spectrum radio
sources: A search for high-redshift radio galaxies}.
\newblock {\em Astron. Astrophys.} {\bf 2014}, {\em 569},~A52,
 doi:10.1051/0004-6361/201423644.

\bibitem[{Williams} {et~al.}(2018){Williams}, {Calistro Rivera}, {Best},
{Hardcastle}, {R{\"o}ttgering}, {Duncan}, {de Gasperin}, {Jarvis}, {Miley},
{Mahony}, {Morabito}, {Nisbet}, {Prandoni}, {Smith}, {Tasse}, and
{White}]{2018MNRAS.475.3429W}
{Williams}, W.L.; {Calistro Rivera}, G.; {Best}, P.N.; {Hardcastle}, M.J.;
{R{\"o}ttgering}, H.J.A.; {Duncan}, K.J.; {de Gasperin}, F.; {Jarvis}, M.J.;
{Miley}, G.K.; {Mahony}, E.K.; et~al.
\newblock {LOFAR-Bo{\"o}tes: Properties of high- and low-excitation radio
galaxies at 0.5 < z < 2.0}.
\newblock {\em Mon. Not. R. Astron. Soc.} {\bf 2018}, {\em 475},~3429--3452,
 doi:10.1093/mnras/sty026.

\bibitem[{Capetti} {et~al.}(2020){Capetti}, {Brienza}, {Baldi},
{Giovannini}, {Morganti}, {Hardcastle}, {Rottgering}, {Brunetti}, {Best}, and
{Miley}]{2020A&A...642A.107C}
{Capetti}, A.; {Brienza}, M.; {Baldi}, R.D.; {Giovannini}, G.; {Morganti}, R.;
{Hardcastle}, M.J.; {Rottgering}, H.J.A.; {Brunetti}, G.F.; {Best}, P.N.;
{Miley}, G.
\newblock {The LOFAR view of FR 0 radio galaxies}.
\newblock {\em Astron. Astrophys.} {\bf 2020}, {\em 642},~A107,
 doi:10.1051/0004-6361/202038671.

\bibitem[{Nakoneczny} {et~al.}(2021){Nakoneczny}, {Bilicki}, {Pollo},
{Asgari}, {Dvornik}, {Erben}, {Giblin}, {Heymans}, {Hildebrandt},
{Kannawadi}, {Kuijken}, {Napolitano}, and {Valentijn}]{2021A&A...649A..81N}
{Nakoneczny}, S.J.; {Bilicki}, M.; {Pollo}, A.; {Asgari}, M.; {Dvornik}, A.;
{Erben}, T.; {Giblin}, B.; {Heymans}, C.; {Hildebrandt}, H.; {Kannawadi}, A.;
et~al.
\newblock {Photometric selection and redshifts for quasars in the Kilo-Degree
Survey Data Release 4}.
\newblock {\em Astron. Astrophys.} {\bf 2021}, {\em 649},~A81,
 doi:10.1051/0004-6361/202039684.

\bibitem[{Wenzl} {et~al.}(2021){Wenzl}, {Schindler}, {Fan}, {Andika},
{Ba{\~n}ados}, {Decarli}, {Jahnke}, {Mazzucchelli}, {Onoue}, {Venemans},
{Walter}, and {Yang}]{2021AJ....162...72W}
{Wenzl}, L.; {Schindler}, J.T.; {Fan}, X.; {Andika}, I.T.; {Ba{\~n}ados}, E.;
{Decarli}, R.; {Jahnke}, K.; {Mazzucchelli}, C.; {Onoue}, M.; {Venemans},
B.P.; et~al.
\newblock {Random Forests as a Viable Method to Select and Discover
High-redshift Quasars}.
\newblock {\em Astron. J.} {\bf 2021}, {\em 162},~72,
 doi:10.3847/1538-3881/ac0254.

\bibitem[{Ma} {et~al.}(2019){Ma}, {Xu}, {Zhu}, {Hu}, {Li}, {Shan}, {Zhu},
{Gu}, {Li}, {Liu}, and {Wu}]{2019ApJS..240...34M}
{Ma}, Z.; {Xu}, H.; {Zhu}, J.; {Hu}, D.; {Li}, W.; {Shan}, C.; {Zhu}, Z.; {Gu},
L.; {Li}, J.; {Liu}, C.; et~al.
\newblock {A Machine Learning Based Morphological Classification of 14,245
Radio AGNs Selected from the Best-Heckman Sample}.
\newblock {\em Astrophys. J.} {\bf 2019}, {\em 240},~34,
 doi:10.3847/1538-4365/aaf9a2.

\bibitem[{Lukic} {et~al.}(2019){Lukic}, {Br{\"u}ggen}, {Mingo}, {Croston},
{Kasieczka}, and {Best}]{2019MNRAS.487.1729L}
{Lukic}, V.; {Br{\"u}ggen}, M.; {Mingo}, B.; {Croston}, J.H.; {Kasieczka}, G.;
{Best}, P.N.
\newblock {Morphological classification of radio galaxies: Capsule networks
versus convolutional neural networks}.
\newblock {\em Mon. Not. R. Astron. Soc.} {\bf 2019}, {\em 487},~1729--1744,
 doi:10.1093/mnras/stz1289.

\bibitem[{Mostert} {et~al.}(2021){Mostert}, {Duncan}, {R{\"o}ttgering},
{Polsterer}, {Best}, {Brienza}, {Br{\"u}ggen}, {Hardcastle}, {Jurlin},
{Mingo}, {Morganti}, {Shimwell}, {Smith}, and
{Williams}]{2021A&A...645A..89M}
{Mostert}, R.I.J.; {Duncan}, K.J.; {R{\"o}ttgering}, H.J.A.; {Polsterer}, K.L.;
{Best}, P.N.; {Brienza}, M.; {Br{\"u}ggen}, M.; {Hardcastle}, M.J.; {Jurlin},
N.; {Mingo}, B.; et~al.
\newblock {Unveiling the rarest morphologies of the LOFAR Two-metre Sky Survey
radio source population with self-organised maps}.
\newblock {\em Astron. Astrophys.} {\bf 2021}, {\em 645},~A89,
 doi:10.1051/0004-6361/202038500.

\bibitem[{Vardoulaki} {et~al.}(2021){Vardoulaki}, {Jim{\'e}nez Andrade},
{Delvecchio}, {Smol{\v{c}}i{\'c}}, {Schinnerer}, {Sargent}, {Gozaliasl},
{Finoguenov}, {Bondi}, {Zamorani}, {Badescu}, {Leslie}, {Ceraj},
{Tisani{\'c}}, {Karim}, {Magnelli}, {Bertoldi}, {Romano-Diaz}, and
{Harrington}]{2021A&A...648A.102V}
{Vardoulaki}, E.; {Jim{\'e}nez Andrade}, E.F.; {Delvecchio}, I.;
{Smol{\v{c}}i{\'c}}, V.; {Schinnerer}, E.; {Sargent}, M.T.; {Gozaliasl}, G.;
{Finoguenov}, A.; {Bondi}, M.; {Zamorani}, G.; et~al.
\newblock {FR-type radio sources at 3 GHz VLA-COSMOS: Relation to physical
properties and large-scale environment}.
\newblock {\em Astron. Astrophys.} {\bf 2021}, {\em 648},~A102,
 doi:10.1051/0004-6361/202039488.

\bibitem[{Burhanudin} {et~al.}(2021){Burhanudin}, {Maund}, {Killestein},
{Ackley}, {Dyer}, {Lyman}, {Ulaczyk}, {Cutter}, {Mong}, {Steeghs},
{Galloway}, {Dhillon}, {O'Brien}, {Ramsay}, {Noysena}, {Kotak}, {Breton},
{Nuttall}, {Pall{\'e}}, {Pollacco}, {Thrane}, {Awiphan}, {Chote}, {Chrimes},
{Daw}, {Duffy}, {Eyles-Ferris}, {Gompertz}, {Heikkil{\"a}}, {Irawati},
{Kennedy}, {Levan}, {Littlefair}, {Makrygianni}, {Mata-S{\'a}nchez},
{Mattila}, {McCormac}, {Mkrtichian}, {Mullaney}, {Sawangwit}, {Stanway},
{Starling}, {Str{\o}m}, {Tooke}, and {Wiersema}]{2021MNRAS.505.4345B}
{Burhanudin}, U.F.; {Maund}, J.R.; {Killestein}, T.; {Ackley}, K.; {Dyer},
M.J.; {Lyman}, J.; {Ulaczyk}, K.; {Cutter}, R.; {Mong}, Y.L.; {Steeghs}, D.;
et~al.
\newblock {Light-curve classification with recurrent neural networks for GOTO:
dealing with imbalanced data}.
\newblock {\em Mon. Not. R. Astron. Soc.} {\bf 2021}, {\em 505},~4345--4361,
 doi:10.1093/mnras/stab1545.

\bibitem[{Saz Parkinson} {et~al.}(2016){Saz Parkinson}, {Xu}, {Yu},
{Salvetti}, {Marelli}, and {Falcone}]{2016ApJ...820....8S}
{Saz Parkinson}, P.M.; {Xu}, H.; {Yu}, P.L.H.; {Salvetti}, D.; {Marelli}, M.;
{Falcone}, A.D.
\newblock {Classification and Ranking of Fermi LAT Gamma-ray Sources from the
3FGL Catalog using Machine Learning Techniques}.
\newblock {\em Astrophys. J.} {\bf 2016}, {\em 820},~8,
 doi:10.3847/0004-637X/820/1/8.

\bibitem[{Chiaro} {et~al.}(2016){Chiaro}, {Salvetti}, {La Mura},
{Giroletti}, {Thompson}, and {Bastieri}]{2016MNRAS.462.3180C}
{Chiaro}, G.; {Salvetti}, D.; {La Mura}, G.; {Giroletti}, M.; {Thompson}, D.J.;
{Bastieri}, D.
\newblock {Blazar flaring patterns (B-FlaP) classifying blazar candidate of
uncertain type in the third Fermi-LAT catalogue by artificial neural
networks}.
\newblock {\em Mon. Not. R. Astron. Soc.} {\bf 2016}, {\em 462},~3180--3195,
 doi:10.1093/mnras/stw1830.

\bibitem[{Xiao} {et~al.}(2020){Xiao}, {Cao}, {Fan}, {Costantin}, {Luo}, and
{Pei}]{2020A&C....3200387X}
{Xiao}, H.B.; {Cao}, H.T.; {Fan}, J.H.; {Costantin}, D.; {Luo}, G.Y.; {Pei},
Z.Y.
\newblock {Efficient Fermi source identification with machine learning
methods}.
\newblock {\em Astron. Comput.} {\bf 2020}, {\em 32},~100387,
 doi:10.1016/j.ascom.2020.100387.

\bibitem[{Wang} {et~al.}(2021){Wang}, {Bai}, {L{\'o}pez-Sanjuan}, {Yuan},
{Wang}, {Liu}, {Sobral}, {Baqui}, {Mart{\'\i}n}, {Galarza}, {Alcaniz},
{Angulo}, {Cenarro}, {Crist{\'o}bal-Hornillos}, {Dupke}, {Ederoclite},
{Hern{\'a}ndez-Monteagudo}, {Mar{\'\i}n-Franch}, {Moles}, {Sodr{\'e}},
{V{\'a}zquez Rami{\'o}}, and {Varela}]{2021arXiv210612787W}
{Wang}, C.; {Bai}, Y.; {L{\'o}pez-Sanjuan}, C.; {Yuan}, H.; {Wang}, S.; {Liu},
J.; {Sobral}, D.; {Baqui}, P.O.; {Mart{\'\i}n}, E.L.; {Galarza}, C.A.;
et~al.
\newblock {J-PLUS: Support Vector Machine Applied to
STAR-GALAXY-QSOClassification}.
\newblock {\em arXiv} {\bf 2021}, arXiv:2106.12787.

\bibitem[{Li} {et~al.}(2021){Li}, {Ni}, {Croft}, {Di Matteo}, {Bird}, and
{Feng}]{2021PNAS..11822038L}
{Li}, Y.; {Ni}, Y.; {Croft}, R.A.C.; {Di Matteo}, T.; {Bird}, S.; {Feng}, Y.
\newblock {AI-assisted superresolution cosmological simulations}.
\newblock {\em Proc. Natl. Acad. Sci. USA} {\bf 2021}, {\em
118},~e2022038118,  doi:10.1073/pnas.2022038118.

\bibitem[{Ball} and {Brunner}(2010)]{2010IJMPD..19.1049B}
{Ball}, N.M.; {Brunner}, R.J.
\newblock {Data Mining and Machine Learning in Astronomy}.
\newblock {\em Int. J. Mod. Phys. D} {\bf 2010}, {\em
19},~1049--1106,  doi:10.1142/S0218271810017160.

\bibitem[{Baron}(2019)]{2019arXiv190407248B}
{Baron}, D.
\newblock {Machine Learning in Astronomy: A practical overview}.
\newblock {\em arXiv} {\bf 2019}, arXiv:1904.07248.

\bibitem[{Goebel} {et~al.}(2018){Goebel}, {Chander}, {Holzinger}, {Lecue},
{Akata}, {Stumpf}, {Kieseberg}, and {Holzinger}]{goebel2018explainable}
{Goebel}, R.; {Chander}, A.; {Holzinger}, K.; {Lecue}, F.; {Akata}, Z.;
{Stumpf}, S.; {Kieseberg}, P.; {Holzinger}, A.
\newblock Explainable ai: The new 42?
\newblock In Proceedings of the International Cross-Domain Conference for Machine Learning and
Knowledge Extraction, Hamburg, Germany, 27--30 August 2018; Springer International Publishing: Cham, Switzerland, 2018; pp.
295--303.

\bibitem[{Roscher} {et~al.}(2020){Roscher}, {Bohn}, {Duarte}, and
{Garcke}]{9007737}
{Roscher}, R.; {Bohn}, B.; {Duarte}, M.F.; {Garcke}, J.
\newblock Explainable Machine Learning for Scientific Insights and Discoveries.
\newblock {\em IEEE Access} {\bf 2020}, {\em 8},~42200--42216, doi:10.1109/ACCESS.2020.2976199.

\bibitem[Shapley(1953)]{Shapley_article}
Shapley, L.S., A Value for n-Person Games.
\newblock In {\em Contributions to the Theory of Games (AM-28), Volume II};
Princeton University Press: Princeton, NJ, USA, 1953; Volume~1, pp. 307--318, doi:10.1515/9781400881970-018.

\bibitem[Molnar(2019)]{molnar2019}
Molnar, C.
\newblock Interpretable Machine Learning. 2019.
\newblock Available online: \url{https://christophm.github.io/interpretable-ml-book/} (accessed~on 4 May 2021).

\bibitem[{Marocco} {et~al.}(2021){Marocco}, {Eisenhardt}, {Fowler},
{Kirkpatrick}, {Meisner}, {Schlafly}, {Stanford}, {Garcia}, {Caselden},
{Cushing}, {Cutri}, {Faherty}, {Gelino}, {Gonzalez}, {Jarrett}, {Koontz},
{Mainzer}, {Marchese}, {Mobasher}, {Schlegel}, {Stern}, {Teplitz}, and
{Wright}]{2021ApJS..253....8M}
{Marocco}, F.; {Eisenhardt}, P.R.M.; {Fowler}, J.W.; {Kirkpatrick}, J.D.;
{Meisner}, A.M.; {Schlafly}, E.F.; {Stanford}, S.A.; {Garcia}, N.;
{Caselden}, D.; {Cushing}, M.C.; et~al.
\newblock {The CatWISE2020 Catalog}.
\newblock {\em Astrophys. J.} {\bf 2021}, {\em 253},~8,
 doi:10.3847/1538-4365/abd805.

\bibitem[{Fernique} {et~al.}(2014){Fernique}, {Boch}, {Donaldson}, {Durand},
{O'Mullane}, {Reinecke}, and {Taylor}]{2014ivoa.spec.0602F}
{Fernique}, P.; {Boch}, T.; {Donaldson}, T.; {Durand}, D.; {O'Mullane}, W.;
{Reinecke}, M.; {Taylor}, M.
\newblock MOC---HEALPix Multi-Order Coverage map Version 1.0.
\newblock 	\emph{arXiv} \textbf{2015}, arXiv:1505.02937.

\bibitem[{Flewelling} {et~al.}(2020){Flewelling}, {Magnier}, {Chambers},
{Heasley}, {Holmberg}, {Huber}, {Sweeney}, {Waters}, {Calamida}, {Casertano},
{Chen}, {Farrow}, {Hasinger}, {Henderson}, {Long}, {Metcalfe}, {Narayan},
{Nieto-Santisteban}, {Norberg}, {Rest}, {Saglia}, {Szalay}, {Thakar},
{Tonry}, {Valenti}, {Werner}, {White}, {Denneau}, {Draper}, {Hodapp},
{Jedicke}, {Kaiser}, {Kudritzki}, {Price}, {Wainscoat}, {Chastel}, {McLean},
{Postman}, and {Shiao}]{2020ApJS..251....7F}
{Flewelling}, H.A.; {Magnier}, E.A.; {Chambers}, K.C.; {Heasley}, J.N.;
{Holmberg}, C.; {Huber}, M.E.; {Sweeney}, W.; {Waters}, C.Z.; {Calamida}, A.;
{Casertano}, S.;  et~al.
\newblock {The Pan-STARRS1 Database and Data Products}.
\newblock {\em Astrophys. J.} {\bf 2020}, {\em 251},~7,
 doi:10.3847/1538-4365/abb82d.

\bibitem[{Bianchi} {et~al.}(2017){Bianchi}, {Shiao}, and
{Thilker}]{2017ApJS..230...24B}
{Bianchi}, L.; {Shiao}, B.; {Thilker}, D.
\newblock {Revised Catalog of GALEX Ultraviolet Sources. I. The All-Sky Survey:
GUVcat\_AIS}.
\newblock {\em Astrophys. J.} {\bf 2017}, {\em 230},~24,
 doi:10.3847/1538-4365/aa7053.

\bibitem[{Intema} {et~al.}(2017){Intema}, {Jagannathan}, {Mooley}, and
{Frail}]{2017A&A...598A..78I}
{Intema}, H.T.; {Jagannathan}, P.; {Mooley}, K.P.; {Frail}, D.A.
\newblock {The GMRT 150 MHz all-sky radio survey. First alternative data
release TGSS ADR1}.
\newblock {\em Astron. Astrophys.} {\bf 2017}, {\em 598},~A78,
 doi:10.1051/0004-6361/201628536.

\bibitem[{Traulsen} {et~al.}(2020){Traulsen}, {Schwope}, {Lamer}, {Ballet},
{Carrera}, {Ceballos}, {Coriat}, {Freyberg}, {Koliopanos}, {Kurpas},
{Michel}, {Motch}, {Page}, {Watson}, and {Webb}]{2020A&A...641A.137T}
{Traulsen}, I.; {Schwope}, A.D.; {Lamer}, G.; {Ballet}, J.; {Carrera}, F.J.;
{Ceballos}, M.T.; {Coriat}, M.; {Freyberg}, M.J.; {Koliopanos}, F.; {Kurpas},
J.;  et~al.
\newblock {The XMM-Newton serendipitous survey. X. The second source catalogue
from overlapping XMM-Newton observations and its long-term variable content}.
\newblock {\em Astron. Astrophys.} {\bf 2020}, {\em 641},~A137,
 doi:10.1051/0004-6361/202037706.

\bibitem[{Cutri} {et~al.}(2003){Cutri}, {Skrutskie}, {van Dyk}, {Beichman},
{Carpenter}, {Chester}, {Cambresy}, {Evans}, {Fowler}, {Gizis}, {Howard},
{Huchra}, {Jarrett}, {Kopan}, {Kirkpatrick}, {Light}, {Marsh}, {McCallon},
{Schneider}, {Stiening}, {Sykes}, {Weinberg}, {Wheaton}, {Wheelock}, and
{Zacarias}]{2003tmc..book.....C}
{Cutri}, R.M.; {Skrutskie}, M.F.; {van Dyk}, S.; {Beichman}, C.A.; {Carpenter},
J.M.; {Chester}, T.; {Cambresy}, L.; {Evans}, T.; {Fowler}, J.; {Gizis}, J.;
 et~al.
\newblock 2MASS All Sky Catalog of Point Sources. 2003. Available online: \url{https://ui.adsabs.harvard.edu/abs/2003tmc..book.....C/abstract}  (accessed~on 29 May 2021).

\bibitem[{Skrutskie} {et~al.}(2006){Skrutskie}, {Cutri}, {Stiening},
{Weinberg}, {Schneider}, {Carpenter}, {Beichman}, {Capps}, {Chester},
{Elias}, {Huchra}, {Liebert}, {Lonsdale}, {Monet}, {Price}, {Seitzer},
{Jarrett}, {Kirkpatrick}, {Gizis}, {Howard}, {Evans}, {Fowler}, {Fullmer},
{Hurt}, {Light}, {Kopan}, {Marsh}, {McCallon}, {Tam}, {Van Dyk}, and
{Wheelock}]{2006AJ....131.1163S}
{Skrutskie}, M.F.; {Cutri}, R.M.; {Stiening}, R.; {Weinberg}, M.D.;
{Schneider}, S.; {Carpenter}, J.M.; {Beichman}, C.; {Capps}, R.; {Chester},
T.; {Elias}, J.; et~al.
\newblock {The Two Micron All Sky Survey (2MASS)}.
\newblock {\em Astron. J.} {\bf 2006}, {\em 131},~1163--1183, doi:10.1086/498708.

\bibitem[{Cutri} {et~al.}(2013){Cutri}, {Wright}, {Conrow}, {Fowler},
{Eisenhardt}, {Grillmair}, {Kirkpatrick}, {Masci}, {McCallon}, {Wheelock},
{Fajardo-Acosta}, {Yan}, {Benford}, {Harbut}, {Jarrett}, {Lake}, {Leisawitz},
{Ressler}, {Stanford}, {Tsai}, {Liu}, {Helou}, {Mainzer}, {Gettings},
{Gonzalez}, {Hoffman}, {Marsh}, {Padgett}, {Skrutskie}, {Beck}, {Papin}, and
{Wittman}]{2013wise.rept....1C}
{Cutri}, R.M.; {Wright}, E.L.; {Conrow}, T.; {Fowler}, J.W.; {Eisenhardt},
P.R.M.; {Grillmair}, C.; {Kirkpatrick}, J.D.; {Masci}, F.; {McCallon}, H.L.;
{Wheelock}, S.L.;  et~al.
Explanatory Supplement to the AllWISE Data Release Products. 2013. Available online: \url{https://ui.adsabs.harvard.edu/abs/2013wise.rept....1C} (accessed~on 29 May 2021).

\bibitem[{Flesch}(2021)]{2021arXiv210512985F}
{Flesch}, E.W.
\newblock {The Million Quasars (Milliquas) v7.2 Catalogue, now with VLASS
associations. The inclusion of SDSS-DR16Q quasars is detailed}.
\newblock {\em arXiv} {\bf 2021}, arXiv:2105.12985.

\bibitem[{Lyke} {et~al.}(2020){Lyke}, {Higley}, {McLane}, {Schurhammer},
{Myers}, {Ross}, {Dawson}, {Chabanier}, {Martini}, {Busca}, {Mas des
Bourboux}, {Salvato}, {Streblyanska}, {Zarrouk}, {Burtin}, {Anderson},
{Bautista}, {Bizyaev}, {Brandt}, {Brinkmann}, {Brownstein}, {Comparat},
{Green}, {de la Macorra}, {Mu{\~n}oz Guti{\'e}rrez}, {Hou}, {Newman},
{Palanque-Delabrouille}, {P{\^a}ris}, {Percival}, {Petitjean}, {Rich},
{Rossi}, {Schneider}, {Smith}, {Vivek}, and {Weaver}]{2020ApJS..250....8L}
{Lyke}, B.W.; {Higley}, A.N.; {McLane}, J.N.; {Schurhammer}, D.P.; {Myers},
A.D.; {Ross}, A.J.; {Dawson}, K.; {Chabanier}, S.; {Martini}, P.; {Busca},
N.G.;  et~al.
\newblock {The Sloan Digital Sky Survey Quasar Catalog: Sixteenth Data
Release}.
\newblock {\em Astrophys. J.} {\bf 2020}, {\em 250},~8,
 doi:10.3847/1538-4365/aba623.

\bibitem[{D'Isanto} {et~al.}(2018){D'Isanto}, {Cavuoti}, {Gieseke}, and
{Polsterer}]{2018A&A...616A..97D}
{D'Isanto}, A.; {Cavuoti}, S.; {Gieseke}, F.; {Polsterer}, K.L.
\newblock {Return of the features. Efficient feature selection and
interpretation for photometric redshifts}.
\newblock {\em Astron. Astrophys.} {\bf 2018}, {\em 616},~A97,
 doi:10.1051/0004-6361/201833103.

\bibitem[{Hildebrandt} {et~al.}(2010){Hildebrandt}, {Arnouts}, {Capak},
{Moustakas}, {Wolf}, {Abdalla}, {Assef}, {Banerji}, {Ben{\'\i}tez},
{Brammer}, {Budav{\'a}ri}, {Carliles}, {Coe}, {Dahlen}, {Feldmann}, {Gerdes},
{Gillis}, {Ilbert}, {Kotulla}, {Lahav}, {Li}, {Miralles}, {Purger},
{Schmidt}, and {Singal}]{2010A&A...523A..31H}
{Hildebrandt}, H.; {Arnouts}, S.; {Capak}, P.; {Moustakas}, L.A.; {Wolf}, C.;
{Abdalla}, F.B.; {Assef}, R.J.; {Banerji}, M.; {Ben{\'\i}tez}, N.; {Brammer},
G.B.;  et~al.
\newblock {PHAT: PHoto-z Accuracy Testing}.
\newblock {\em Astron. Astrophys.} {\bf 2010}, {\em 523},~A31,
 doi:10.1051/0004-6361/201014885.

\bibitem[{Bernstein} and {Huterer}(2010)]{2010MNRAS.401.1399B}
{Bernstein}, G.; {Huterer}, D.
\newblock {Catastrophic photometric redshift errors: Weak-lensing survey
requirements}.
\newblock {\em Mon. Not. R. Astron. Soc.} {\bf 2010}, {\em 401},~1399--1408,
 doi:10.1111/j.1365-2966.2009.15748.x.

\bibitem[{Henghes} {et~al.}(2021){Henghes}, {Pettitt}, {Thiyagalingam},
{Hey}, and {Lahav}]{2021arXiv210401875H}
{Henghes}, B.; {Pettitt}, C.; {Thiyagalingam}, J.; {Hey}, T.; {Lahav}, O.
\newblock {Benchmarking and scalability of machine-learning methods for
photometric redshift estimation}.
\newblock {\em Mon. Not. R. Astron. Soc.} {\bf 2021}, {\em 505},~4847--4856,
 doi:10.1093/mnras/stab1513.

\bibitem[{Brescia} {et~al.}(2021){Brescia}, {Cavuoti}, {Razim}, {Amaro},
{Riccio}, and {Longo}]{10.3389/fspas.2021.658229}
{Brescia}, M.; {Cavuoti}, S.; {Razim}, O.; {Amaro}, V.; {Riccio}, G.; {Longo},
G.
\newblock Photometric Redshifts With Machine Learning, Lights and Shadows on a
Complex Data Science Use Case.
\newblock {\em Front. Astron. Space Sci.} {\bf 2021}, {\em
8},~70, doi:10.3389/fspas.2021.658229.

\bibitem[Ali(2020)]{PyCaret}
Ali, M. PyCaret: An Open Source, Low-Code Machine Learning Library in Python. PyCaret Version 2.3. 2020.
 Available online: \url{https://www.pycaret.org} (accessed~on 23 October 2021).

\bibitem[Chattopadhyay(2017)]{ChattopadhyayData}
Chattopadhyay, A.K. Incomplete Data in Astrostatistics.
\newblock In {\em Wiley StatsRef: Statistics Reference Online}; American Cancer
Society: Atlanta, GA, USA, 2017; pp. 1--12, https://doi.org/10.1002/9781118445112.stat07942.

\bibitem[Bilogur {et~al.}(2021)Bilogur, samuelbr, Beutner, Fandango,
Everson, Chacreton, Abahurire, Mavroforakis, Cruz, Mahlke, Sergiu, Brugman,
Gates, Todd, and Kamvar]{aleksey_bilogur_2021_5068743}
Bilogur, A.; Samuelbr; Beutner, V.; Fandango, A.; Everson, B.; Chacreton, D.;
Abahurire, E.J.; Mavroforakis, H.; Cruz, J.S.; Mahlke, M.; et al.
\newblock ResidentMario/missingno: 0.5.0 maintenance release. \emph{Zenodo} \textbf{2021}, doi:10.5281/zenodo.5068743.

\bibitem[Kursa and Rudnicki(2010)]{JSSv036i11}
Kursa, M.B.; Rudnicki, W.R.
\newblock Feature Selection with the Boruta Package.
\newblock {\em J. Stat. Softw. Artic.} {\bf 2010}, {\em
36},~1--13, doi:10.18637/jss.v036.i11.

\bibitem[Yeo and Johnson(2000)]{10.1093/biomet/87.4.954}
Yeo, I.; Johnson, R.A.
\newblock {A new family of power transformations to improve normality or
symmetry}.
\newblock {\em Biometrika} {\bf 2000}, {\em 87},~954--959,
 doi:10.1093/biomet/87.4.954.

\bibitem[Geurts {et~al.}(2006)Geurts, Ernst, and Wehenkel]{Geurts2006}
Geurts, P.; Ernst, D.; Wehenkel, L.
\newblock Extremely randomized trees.
\newblock {\em Mach. Learn.} {\bf 2006}, {\em 63},~3--42, doi:10.1007/s10994-006-6226-1.

\bibitem[Dorogush {et~al.}(2017)Dorogush, Gulin, Gusev, Kazeev,
Prokhorenkova, and Vorobev]{DBLP:journals/corr/DorogushGGKPV17}
Dorogush, A.V.; Gulin, A.; Gusev, G.; Kazeev, N.; Prokhorenkova, L.O.; Vorobev,
A.
\newblock Fighting biases with dynamic boosting.
\newblock {\em arXiv} {\bf 2017}, {\em arXiv:1706.09516}.

\bibitem[Dorogush {et~al.}(2018)Dorogush, Ershov, and
Gulin]{DBLP:journals/corr/abs-1810-11363}
Dorogush, A.V.; Ershov, V.; Gulin, A.
\newblock CatBoost: Gradient boosting with categorical features support.
\newblock {\em arXiv} {\bf 2018}, {\em arXiv:1810.11363}.

\bibitem[Ke {et~al.}(2017)Ke, Meng, Finley, Wang, Chen, Ma, Ye, and
Liu]{NIPS2017_6449f44a}
Ke, G.; Meng, Q.; Finley, T.; Wang, T.; Chen, W.; Ma, W.; Ye, Q.; Liu, T.Y.
\newblock LightGBM: A Highly Efficient Gradient Boosting Decision Tree.
\newblock In \emph{Advances in Neural Information Processing Systems}; Guyon, I.,
Luxburg, U.V., Bengio, S., Wallach, H., Fergus, R., Vishwanathan, S.,
Garnett, R., Eds.; Curran Associates, Inc.: Red Hook, NY, USA, 2017; Volume~30.

\bibitem[Chen and Guestrin(2016)]{Chen:2016:XST:2939672.2939785}
Chen, T.; Guestrin, C.
\newblock {XGBoost}: A Scalable Tree Boosting System.
\newblock In Proceedings of the KDD '16: The 22nd ACM SIGKDD International Conference on Knowledge Discovery and Data Mining, San Francisco, CA, USA, 13--17 August 2016; ACM: New York, NY, USA, 2016; pp. 785--794, doi:10.1145/2939672.2939785.

\bibitem[Breiman(2001)]{Breiman2001}
Breiman, L.
\newblock Random Forests.
\newblock {\em Mach. Learn.} {\bf 2001}, {\em 45},~5--32, doi:10.1023/A:1010933404324.

\bibitem[{Ananna} {et~al.}(2017){Ananna}, {Salvato}, {LaMassa}, {Urry},
{Cappelluti}, {Cardamone}, {Civano}, {Farrah}, {Gilfanov}, {Glikman},
{Hamilton}, {Kirkpatrick}, {Lanzuisi}, {Marchesi}, {Merloni}, {Nandra},
{Natarajan}, {Richards}, and {Timlin}]{2017ApJ...850...66A}
{Ananna}, T.T.; {Salvato}, M.; {LaMassa}, S.; {Urry}, C.M.; {Cappelluti}, N.;
{Cardamone}, C.; {Civano}, F.; {Farrah}, D.; {Gilfanov}, M.; {Glikman}, E.;
 et~al.
\newblock {AGN Populations in Large-volume X-Ray Surveys: Photometric Redshifts
and Population Types Found in the Stripe 82X Survey}.
\newblock {\em Astrophys. J.} {\bf 2017}, {\em 850},~66,
 doi:10.3847/1538-4357/aa937d.

\bibitem[{Hodge} {et~al.}(2011){Hodge}, {Becker}, {White}, {Richards}, and
{Zeimann}]{2011AJ....142....3H}
{Hodge}, J.A.; {Becker}, R.H.; {White}, R.L.; {Richards}, G.T.; {Zeimann}, G.R.
\newblock {High-resolution Very Large Array Imaging of Sloan Digital Sky Survey
Stripe 82 at 1.4 GHz}.
\newblock {\em Astron. J.} {\bf 2011}, {\em 142},~3,
 doi:10.1088/0004-6256/142/1/3.

\bibitem[{Curran} {et~al.}(2021){Curran}, {Moss}, and
{Perrott}]{2021MNRAS.503.2639C}
{Curran}, S.J.; {Moss}, J.P.; {Perrott}, Y.C.
\newblock {QSO photometric redshifts using machine learning and neural
networks}.
\newblock {\em Mon. Not. R. Astron. Soc.} {\bf 2021}, {\em 503},~2639--2650,
 doi:10.1093/mnras/stab485.

\bibitem[Lundberg and Lee(2017)]{NIPS2017_7062}
Lundberg, S.M.; Lee, S.I.
\newblock A Unified Approach to Interpreting Model Predictions. In {\em
Advances in Neural Information Processing Systems 30}; Guyon, I., Luxburg,
U.V., Bengio, S., Wallach, H., Fergus, R., Vishwanathan, S., Garnett, R.,
Eds.; Curran Associates, Inc.: Red Hook, NY, USA, 2017; pp. 4765--4774.

\bibitem[Lundberg {et~al.}(2020)Lundberg, Erion, Chen, DeGrave, Prutkin,
Nair, Katz, Himmelfarb, Bansal, and Lee]{lundberg2020local2global}
Lundberg, S.M.; Erion, G.; Chen, H.; DeGrave, A.; Prutkin, J.M.; Nair, B.;
Katz, R.; Himmelfarb, J.; Bansal, N.; Lee, S.I.
\newblock From local explanations to global understanding with explainable AI
for trees.
\newblock {\em Nat. Mach. Intell.} {\bf 2020}, {\em 2},~2522--5839.

\bibitem[{Turner} {et~al.}(2020){Turner}, {Drouart}, {Seymour}, and
{Shabala}]{2020MNRAS.499.3660T}
{Turner}, R.J.; {Drouart}, G.; {Seymour}, N.; {Shabala}, S.S.
\newblock {RAiSERed: Radio continuum redshifts for lobed active galactic
nuclei}.
\newblock {\em Mon. Not. R. Astron. Soc.} {\bf 2020}, {\em 499},~3660--3672,
 doi:10.1093/mnras/staa3067.

\bibitem[{Zaja{\v{c}}ek} {et~al.}(2019){Zaja{\v{c}}ek}, {Busch},
{Valencia-S.}, {Eckart}, {Britzen}, {Fuhrmann}, {Schneeloch}, {Fazeli},
{Harrington}, and {Zensus}]{2019A&A...630A..83Z}
{Zaja{\v{c}}ek}, M.; {Busch}, G.; {Valencia-S.}, M.; {Eckart}, A.; {Britzen},
S.; {Fuhrmann}, L.; {Schneeloch}, J.; {Fazeli}, N.; {Harrington}, K.C.;
{Zensus}, J.A.
\newblock {Radio spectral index distribution of SDSS-FIRST sources across
optical diagnostic diagrams}.
\newblock {\em Astron. Astrophys.} {\bf 2019}, {\em 630},~A83,
 doi:10.1051/0004-6361/201833388.

\bibitem[{Laor} and {Behar}(2008)]{2008MNRAS.390..847L}
{Laor}, A.; {Behar}, E.
\newblock {On the origin of radio emission in radio-quiet quasars}.
\newblock {\em Mon. Not. R. Astron. Soc.} {\bf 2008}, {\em 390},~847--862,
 doi:10.1111/j.1365-2966.2008.13806.x.

\bibitem[{Laor} {et~al.}(2019){Laor}, {Baldi}, and
{Behar}]{2019MNRAS.482.5513L}
{Laor}, A.; {Baldi}, R.D.; {Behar}, E.
\newblock {What drives the radio slopes in radio-quiet quasars?}
\newblock {\em Mon. Not. R. Astron. Soc.} {\bf 2019}, {\em 482},~5513--5523,
 doi:10.1093/mnras/sty3098.

\bibitem[{McKean} {et~al.}(2021){McKean}, {Luichies}, {Drabent},
{G{\"u}rkan}, {Hartley}, {Lafontaine}, {Prandoni}, {R{\"o}ttgering},
{Shimwell}, {Stacey}, and {Tasse}]{2021MNRAS.505L..36M}
{McKean}, J.P.; {Luichies}, R.; {Drabent}, A.; {G{\"u}rkan}, G.; {Hartley}, P.;
{Lafontaine}, A.; {Prandoni}, I.; {R{\"o}ttgering}, H.J.A.; {Shimwell}, T.W.;
{Stacey}, H.R.;  et~al.
\newblock {Gravitational lensing in LoTSS DR2: Extremely faint 144-MHz radio
emission from two highly magnified quasars}.
\newblock {\em Mon. Not. R. Astron. Soc.} {\bf 2021}, {\em 505},~L36--L40,
 doi:10.1093/mnrasl/slab033.

\bibitem[{Driver} and {Robotham}(2010)]{2010MNRAS.407.2131D}
{Driver}, S.P.; {Robotham}, A.S.G.
\newblock {Quantifying cosmic variance}.
\newblock {\em Mon. Not. R. Astron. Soc.} {\bf 2010}, {\em 407},~2131--2140,
 doi:10.1111/j.1365-2966.2010.17028.x.

\bibitem[{Wolf} {et~al.}(2004){Wolf}, {Meisenheimer}, {Kleinheinrich},
{Borch}, {Dye}, {Gray}, {Wisotzki}, {Bell}, {Rix}, {Cimatti}, {Hasinger}, and
{Szokoly}]{2004A&A...421..913W}
{Wolf}, C.; {Meisenheimer}, K.; {Kleinheinrich}, M.; {Borch}, A.; {Dye}, S.;
{Gray}, M.; {Wisotzki}, L.; {Bell}, E.F.; {Rix}, H.W.; {Cimatti}, A.;
 et~al.
\newblock {A catalogue of the Chandra Deep Field South with multi-colour
classification and photometric redshifts from COMBO-17}.
\newblock {\em Astron. Astrophys.} {\bf 2004}, {\em 421},~913--936,
 doi:10.1051/0004-6361:20040525.

\bibitem[{Salvato} {et~al.}(2009){Salvato}, {Hasinger}, {Ilbert},
{Zamorani}, {Brusa}, {Scoville}, {Rau}, {Capak}, {Arnouts}, {Aussel},
{Bolzonella}, {Buongiorno}, {Cappelluti}, {Caputi}, {Civano}, {Cook},
{Elvis}, {Gilli}, {Jahnke}, {Kartaltepe}, {Impey}, {Lamareille}, {Le Floc'h},
{Lilly}, {Mainieri}, {McCarthy}, {McCracken}, {Mignoli}, {Mobasher},
{Murayama}, {Sasaki}, {Sanders}, {Schiminovich}, {Shioya}, {Shopbell},
{Silverman}, {Smol{\v{c}}i{\'c}}, {Surace}, {Taniguchi}, {Thompson}, {Trump},
{Urry}, and {Zamojski}]{2009ApJ...690.1250S}
{Salvato}, M.; {Hasinger}, G.; {Ilbert}, O.; {Zamorani}, G.; {Brusa}, M.;
{Scoville}, N.Z.; {Rau}, A.; {Capak}, P.; {Arnouts}, S.; {Aussel}, H.;
 et~al.
\newblock {Photometric Redshift and Classification for the XMM-COSMOS Sources}.
\newblock {\em Astrophys. J.} {\bf 2009}, {\em 690},~1250--1263,
 doi:10.1088/0004-637X/690/2/1250.

\bibitem[{Matute} {et~al.}(2012){Matute}, {M{\'a}rquez}, {Masegosa},
{Husillos}, {del Olmo}, {Perea}, {Alfaro}, {Fern{\'a}ndez-Soto}, {Moles},
{Aguerri}, {Aparicio-Villegas}, {Ben{\'\i}tez}, {Broadhurst}, {Cabrera-Cano},
{Castander}, {Cepa}, {Cervi{\~n}o}, {Crist{\'o}bal-Hornillos}, {Infante},
{Gonz{\'a}lez Delgado}, {Mart{\'\i}nez}, {Molino}, {Prada}, and
{Quintana}]{2012A&A...542A..20M}
{Matute}, I.; {M{\'a}rquez}, I.; {Masegosa}, J.; {Husillos}, C.; {del Olmo},
A.; {Perea}, J.; {Alfaro}, E.J.; {Fern{\'a}ndez-Soto}, A.; {Moles}, M.;
{Aguerri}, J.A.L.; et~al.
\newblock {Quasi-stellar objects in the ALHAMBRA survey. I. Photometric
redshift accuracy based on 23 optical-NIR filter photometry}.
\newblock {\em Astron. Astrophys.} {\bf 2012}, {\em 542},~A20,
 doi:10.1051/0004-6361/201118111.

\bibitem[{van Haarlem} {et~al.}(2013)]{2013A&A...556A...2V}
{van Haarlem}, M.P.; {Wise}, M.W.; {Gunst}, A.W.; {Heald}, G.; {McKean}, J.P.;
{Hessels}, J.W.T.; {de Bruyn}, A.G.; {Nijboer}, R.; {Swinbank}, J.;
{Fallows}, R.;  et~al.
\newblock {LOFAR: The LOw-Frequency ARray}.
\newblock {\em Astron. Astrophys.} {\bf 2013}, {\em 556},~A2,
 doi:10.1051/0004-6361/201220873.

\bibitem[{Ochsenbein} {et~al.}(2000){Ochsenbein}, {Bauer}, and
{Marcout}]{vizier}
{Ochsenbein}, F.; {Bauer}, P.; {Marcout}, J.
\newblock {The VizieR database of astronomical catalogues}.
\newblock {\em Astron. Astrophys.} {\bf 2000}, {\em 143},~23--32,
 doi:10.1051/aas:2000169.

\bibitem[{Astropy Collaboration} {et~al.}(2013){Astropy Collaboration},
{Robitaille}, {Tollerud}, {Greenfield}, {Droettboom}, {Bray}, {Aldcroft},
{Davis}, {Ginsburg}, {Price-Whelan}, {Kerzendorf}, {Conley}, {Crighton},
{Barbary}, {Muna}, {Ferguson}, {Grollier}, {Parikh}, {Nair}, {Unther},
{Deil}, {Woillez}, {Conseil}, {Kramer}, {Turner}, {Singer}, {Fox}, {Weaver},
{Zabalza}, {Edwards}, {Azalee Bostroem}, {Burke}, {Casey}, {Crawford},
{Dencheva}, {Ely}, {Jenness}, {Labrie}, {Lim}, {Pierfederici}, {Pontzen},
{Ptak}, {Refsdal}, {Servillat}, and {Streicher}]{astropy:2013}
{Astropy Collaboration}; {Robitaille}, T.P.; {Tollerud}, E.J.; {Greenfield},
P.; {Droettboom}, M.; {Bray}, E.; {Aldcroft}, T.; {Davis}, M.; {Ginsburg},
A.; {Price-Whelan}, A.M.;  et~al.
\newblock {Astropy: A community Python package for astronomy}.
\newblock {\em Astron. Astrophys.} {\bf 2013}, {\em 558},~A33,
 doi:10.1051/0004-6361/201322068.

\bibitem[{Astropy Collaboration} {et~al.}(2018){Astropy Collaboration},
{Price-Whelan}, {Sip{\H{o}}cz}, {G{\"u}nther}, {Lim}, {Crawford}, {Conseil},
{Shupe}, {Craig}, {Dencheva}, {Ginsburg}, {Vand erPlas}, {Bradley},
{P{\'e}rez-Su{\'a}rez}, {de Val-Borro}, {Aldcroft}, {Cruz}, {Robitaille},
{Tollerud}, {Ardelean}, {Babej}, {Bach}, {Bachetti}, {Bakanov}, {Bamford},
{Barentsen}, {Barmby}, {Baumbach}, {Berry}, {Biscani}, {Boquien}, {Bostroem},
{Bouma}, {Brammer}, {Bray}, {Breytenbach}, {Buddelmeijer}, {Burke},
{Calderone}, {Cano Rodr{\'\i}guez}, {Cara}, {Cardoso}, {Cheedella}, {Copin},
{Corrales}, {Crichton}, {D'Avella}, {Deil}, {Depagne}, {Dietrich}, {Donath},
{Droettboom}, {Earl}, {Erben}, {Fabbro}, {Ferreira}, {Finethy}, {Fox},
{Garrison}, {Gibbons}, {Goldstein}, {Gommers}, {Greco}, {Greenfield},
{Groener}, {Grollier}, {Hagen}, {Hirst}, {Homeier}, {Horton}, {Hosseinzadeh},
{Hu}, {Hunkeler}, {Ivezi{\'c}}, {Jain}, {Jenness}, {Kanarek}, {Kendrew},
{Kern}, {Kerzendorf}, {Khvalko}, {King}, {Kirkby}, {Kulkarni}, {Kumar},
{Lee}, {Lenz}, {Littlefair}, {Ma}, {Macleod}, {Mastropietro}, {McCully},
{Montagnac}, {Morris}, {Mueller}, {Mumford}, {Muna}, {Murphy}, {Nelson},
{Nguyen}, {Ninan}, {N{\"o}the}, {Ogaz}, {Oh}, {Parejko}, {Parley}, {Pascual},
{Patil}, {Patil}, {Plunkett}, {Prochaska}, {Rastogi}, {Reddy Janga},
{Sabater}, {Sakurikar}, {Seifert}, {Sherbert}, {Sherwood-Taylor}, {Shih},
{Sick}, {Silbiger}, {Singanamalla}, {Singer}, {Sladen}, {Sooley},
{Sornarajah}, {Streicher}, {Teuben}, {Thomas}, {Tremblay}, {Turner},
{Terr{\'o}n}, {van Kerkwijk}, {de la Vega}, {Watkins}, {Weaver}, {Whitmore},
{Woillez}, {Zabalza}, and {Astropy Contributors}]{astropy:2018}
{Astropy Collaboration}; {Price-Whelan}, A.M.; {Sip{\H{o}}cz}, B.M.;
{G{\"u}nther}, H.M.; {Lim}, P.L.; {Crawford}, S.M.; {Conseil}, S.; {Shupe},
D.L.; {Craig}, M.W.; {Dencheva}, N.; et~al.
\newblock {The Astropy Project: Building an Open-science Project and Status of
the v2.0 Core Package}.
\newblock {\em Astron. J.} {\bf 2018}, {\em 156},~123,
 doi:10.3847/1538-3881/aabc4f.

\bibitem[{Taylor}(2005)]{2005ASPC..347...29T}
{Taylor}, M.B.
\newblock {TOPCAT \& STIL: Starlink Table/VOTable Processing Software}.
\newblock In Proceedings of the Astronomical Data Analysis Software and Systems XIV, Pasadena, CA, USA, 24--27 October 2004; {Shopbell}, P.,
{Britton}, M., {Ebert}, R., Eds.; Astronomical Society
of the Pacific Conference Series; Astronomical Society of the Pacific: San Francisco, CA, USA, 2005; Volume 347, p.~29.

\end{thebibliography}
\end{document}